\documentclass{aa}
\usepackage[breaklinks]{hyperref}
\usepackage{orcidlink} 
\usepackage{etoolbox}
\usepackage{graphicx}
\usepackage{natbib}
\usepackage{txfonts}

\usepackage{xcolor}
\usepackage[normalem]{ulem}
\newcommand{\teff}{\ensuremath{T_\mathrm{eff}}}

\bibpunct{(}{)}{;}{a}{}{,}

\begin{document}

\title{TOPoS VII. Age--metallicity relation in the Galactic halo and assembly of the Milky Way}
\titlerunning{TOPoS VII. Age--metallicity in the Galactic halo}
\author{
P.~Bonifacio \inst{1}\orcidlink{0000-0002-1014-0635} \and
Y.~Lebreton \inst {1,2}\orcidlink{0000-0002-4834-2144}\and
I.~Koutsouridou \inst{3}\orcidlink{0000-0002-3524-7172} \and
Y.~P.~Zou \inst{4,5}\orcidlink{0000-0002-5230-8010} \and
D.~Romano \inst{5}\orcidlink{0000-0002-0845-6171}  \and
S.~Salvadori \inst{3} \orcidlink{0000-0001-7298-2478} \and 
L.~Sbordone \inst{6a\orcidlink{0000-0002-2285-8708}} \and
L.~Monaco \inst{7}\orcidlink{0000-0002-3148-9836} \and
E.~Caffau \inst{1}\orcidlink{0000-0001-6011-6134} \and
M.~Spite \inst{1}\orcidlink{0000-0002-2795-3421} \and
P. Fran\c{c}ois \inst{8,9}\orcidlink{0000-0001-8698-2217}
}
\institute{LIRA, Observatoire de Paris, Universit\'e PSL, Sorbonne Universit\'e, Universit\'e Paris Cit\'e, CY Cergy Paris Universit\'e, CNRS, 92190 Meudon, France\relax\label{lira}
\and
{Univ Rennes, CNRS, IPR (Institut de Physique de Rennes) - UMR 6251, F-35000 Rennes, France}
\and
Dipartimento di Fisica e Astronomia, Università degli Studi di Firenze, Via G. Sansone 1, 50019, Sesto Fiorentino, Italy; INAF/Osservatorio Astrofisico di Arcetri, Largo E. Fermi 5, I-50125, Firenze, Italy
\and
Department of Physics and Astronomy, University of Bologna, Via Gobetti 93/2, 40129 Bologna, Italy
\and
INAF, Astrophysics and Space Science Observatory, Via Gobetti 93/3, 40129 Bologna, Italy
\and
European Southern Observatory, Casilla 19001, Santiago, Chile
\and
Universidad Andres Bello, Facultad de Ciencias Exactas, Departamento de F\'isica y Astronom\'ia – Instituto de Astrof\'isica, Autopista Concepci\'on-Talcahuano 7100, Talcahuano, Chile
\and
LIRA, Observatoire de Paris, Universit\'e PSL, Sorbonne Universit\'e, Universit\'e Paris Cit\'e, CY Cergy Paris Universit\'e, CNRS, 75014 Paris, France
\and
UPJV, Universit\'e de Picardie Jules Verne, 33 rue St Leu, 80080 Amiens, France}
\date{Received ; accepted }
\abstract
{One 
technique for determining stellar ages is to compare the position of a star in the Hertzsprung-Russell
diagram to theoretical stellar evolutionary tracks. The sub-giant evolutionary stage
is the one that is most sensitive to age and allows the most precise evolutionary age estimates.} %leave it empty if necessary  
{The TOPoS sample of stars with metallicities derived from low-resolution Sloan Digital Sky Survey spectra contains a large subset
of sub-giant stars with precise parallaxes from the Gaia mission, for which evolutionary ages
can be determined. Our aim is to use this stellar sample to investigate the age--metallicity relation in the 
Galactic halo.}
{We use the Bayesian inference code SPInS and theoretical BaSTI stellar evolutionary tracks
to determine the ages for  TOPoS stars. 
}
%using their metallicities and
%positions in the HR diagram.}
{There is a clear increase in metallicity with decreasing age, albeit with a considerable scatter
at any given age. At ages larger than 8 Ga, the scatter is so large that in fact, over this range, 
age and metallicity appear to be uncorrelated. 
At any given age, the metallicity distribution is multi-modal, with up to three distinct peaks.
These peaks trace three age--metallicity relations that we tentatively identify with the halo, thick-disc, and thin-disc.
}
{Our data demonstrate the important role of mergers in the evolution of the Galaxy, up to 
8 Ga ago. In more recent times, the spread
in metallicity drops. One 
possibility is that the major merger Gaia-Sausage-Enceladus
may have perturbed the galaxies in the Milky Way vicinity in
such a way as to decrease the merger rate. 
Chemical evolution models and cosmological models of the Local Group both support the importance
of mergers in the early evolution of the Milky Way. Larger, unbiased samples, or at least with well- understood
biases, of stars with accurate ages are required for a quantitative comparison between models and data.}
\maketitle

\section{Introduction}

In a previous paper of the series \citep{toposVI},
we exploited the stars that were used to select candidate
extremely metal-poor stars
for high resolution spectroscopic observations
of the TOPoS project \citep{toposI} in order to derive
the metallicity distribution function of the Galactic halo.
In this paper, we want to exploit a subset of that sample
in order to investigate the age--metallicity relation (AMR) in the Galactic halo. 

The metallicity of the gas in galaxies is related to the stellar mass and star formation rate (SFR)
of the galaxy.
This relation is smooth and  can be modelled by a second degree
polynomial \citep[equation 2 of][]{Mannucci2010}.
For galaxies with high stellar masses (above about $10^{10}$ $\rm M_\odot$)
the gas metallicity is essentially independent of the SFR, but for lower masses 
the metallicity decreases with increasing SFR \citep{Ellison2008,Mannucci2010}. 
This is because
a high SFR is generally driven by infalling gas, but this gas
also dilutes the metals in the galaxy gas phase, thus leading to a lower
gas phase metallicity. 
\citet{Lilly2013} developed a model of galaxy evolution
that naturally implies that  gas metallicity is a function of stellar
mass and SFR. This function is capable of fitting the data of \citet{Mannucci2010},
thus providing some theoretical support to the empirical relation.
Their model also considers the effects of outflows
that are proportional to the SFR; however, they conclude that
the fraction of mass in gas to the mass in stars does not
depend on the outflows but only on the efficiency of star
formation and on the SFR per unit stellar mass.

There is an exception to this general rule and that
is the case for the smaller galaxy in a minor merger \citep{MD2008}.

In this case the gas raising the SFR comes from the larger
galaxy that, before the merger starts, has higher metallicity. 
Thus, the infalling gas is pre-enriched and the smaller galaxies
show, on average, a metallicity that is about 0.2\,dex higher
than an isolated galaxy of the same stellar mass. 

The age--metallicity relation is a property of a galaxy that depends
on its SFR and on the evolution of its mass, and therefore its
merger history.   The models discussed above do not consider explicitly merging and assume
that the mass increase of a galaxy is only due to the inflow of gas and dark matter from the
environment of the galaxy. The age--metallicity relation is an output of models of galactic evolution and comparison
of models to observations provides insight into the galactic history. 
Our aim is to provide a set of stars against which various models can be tested. 

\section{Target selection}

\begin{figure}
\resizebox{7.7cm}{!}{\includegraphics{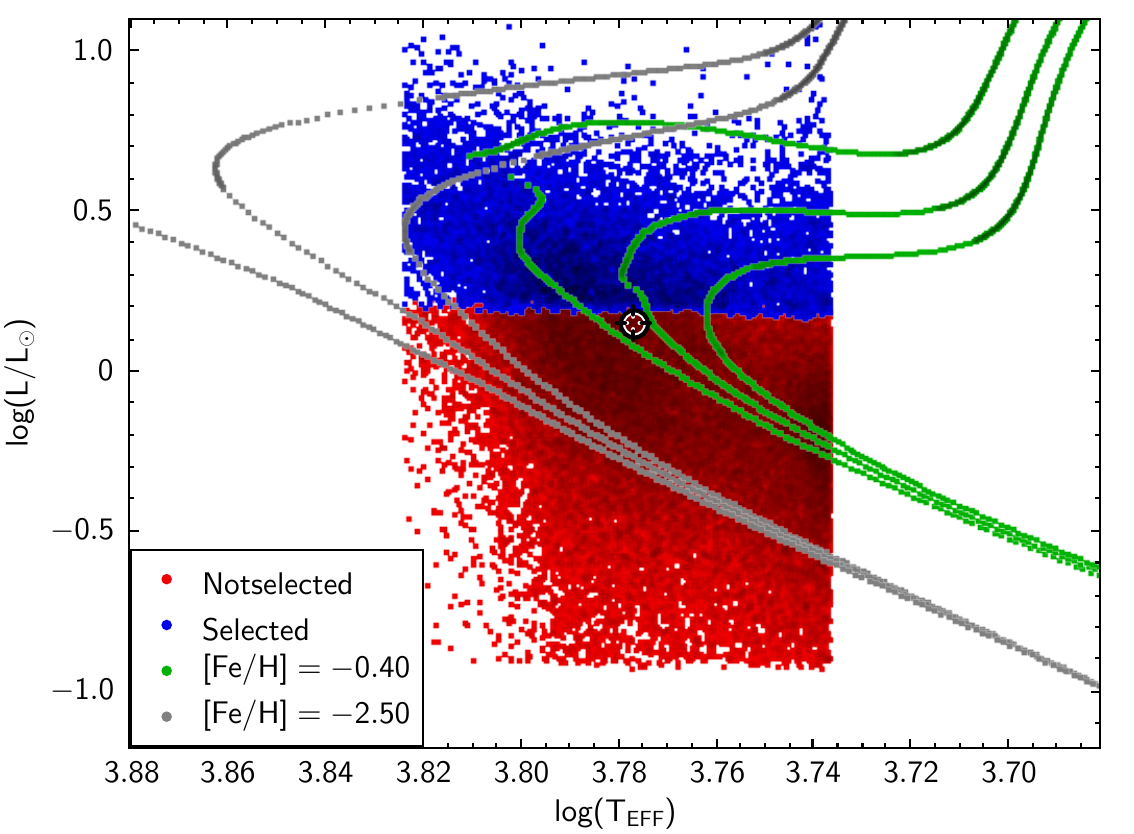}}
\caption{Colour magnitude diagram of the TOPoS stars with
those we analyse as SGs in blue, the others in red. To guide the eye
we superimpose three BaSTI  \citep{pietrinferni2021} isochrones with ages
4, 8, and 12\,Ga and [Fe/H]=--0.4 (green) and --2.5 (grey).}
\label{fig:select}
\end{figure}

We decided to focus on the sub-giant (SG) 
stars in the 'good parallax'
sample of \citet{toposVI}, because this is the evolutionary
stage that is most sensitive to age.
The selection of potential SG stars was
done on the absolute magnitude, as $G_{\rm 0,abs} < 4.2$.
The selection results in a sample of 14\,682
unique stars, where {  abundances from multiple observations of the same star  have been averaged }as described
in \citet{toposVI}; the selection is displayed in Fig.\,\ref{fig:select}. It is clear that for the more
metal-rich stars in our selection we also include turn-off and main sequence stars;
as we go to more metal-poor stars the selection is increasingly dominated by turn-off and 
SG stars. The temperature cut is such that we do not include
young metal-poor turn-off stars if they exist. However, we would
capture young metal-poor SGs. As discussed in \citep{bonifacio2012} and \citep{toposI}, the
blue colour cut $(g-z)_0 \ge 0.18$, which appears in Fig.\,\ref{fig:select} as a temperature cut,
was made to exclude most of the white dwarfs. One should also
keep in mind that the analysis of Sloan Digital Sky Survey (SDSS) spectra for stars of low metallicity warmer than 6500\,K
is extremely challenging due to the weakness of the lines. With this selection, the majority of the stars are SGs, regardless of  their age or metallicity. When we discuss the AMR, we return to the question of whether this selection may introduce any bias.

In Appendix \ref{ap_lum} we detail how we estimate 
luminosities and associated errors from the data in
the catalogue of \citet{toposVI}.
In order to limit our sample to stars of good enough quality, we
only select stars with an error in the logarithmic luminosity less than 0.1.
This results in a sample of 8\,737 stars.
{ 
One could suspect that the cut in luminosity error would greatly
favour nearby stars, thus leaving the more distant stars under-sampled. 
If we focus on the halo stars, which are typically more distant, we note that
in the sample of 14\,682 stars, 29\% are halo stars; however, in the
sample of 8\,737 stars, 22\% are halo. Therefore, the halo is indeed
slightly under-represented in our smaller sample.
}

\section{Determining ages using SPInS}
We used the logarithm base 10 of effective temperatures and luminosities, as well as 
metallicities, 
from \citet{toposVI} as input to SPInS \citep{LebretonReese2020,spins2020, C2024}
to derive ages for our sample of stars. 
The choice of different projection planes does not substantially change the results, 
as shown, for example, by \citet[][figure 5]{Bonifacio2024} and by the generally good agreement
with the results of \citet[][see Sec.\,\ref{sec:ngc6397} and Sec.\,\ref{comp_c24} ]{C2024}, who used absolute magnitudes, colours, and metallicities. 
{  
We assumed an error of 0.018 in the logarithm of effective temperature
that corresponds to  about 250\,K for \teff = 5900\,K.
The uncertainties in the correspondence 
between colours and effective temperatures are those 
implied in the determination of 
effective temperatures} by
\citet[][see there for details]{toposVI}.  
%As error on the logarithm of the effective temperature we assumed 0.018, 
For our sample,
the age error does not correlate with errors in the
effective temperature,  luminosity, or [Fe/H]. Instead, it strongly correlates with
the luminosity: the higher the luminosity, the smaller the error.

As  set of stellar evolutionary tracks we chose the $\alpha$-element enriched ($\rm [\alpha/Fe]=+0.4$)  BaSTI ones \citep{pietrinferni2021},
which take into account atomic diffusion, 
and we imposed a flat prior {  on} ages between 0 and 13.8 Ga,
which prevents finding ages older than the age of the Universe
\citep{planck}.
To check the impact of these assumptions, we also determined ages 
with no priors, both 
with BaSTI stellar models  based on a solar-scaled mixture \citep{hidalgo2018} and 
with $\alpha$-enhanced models.
This test showed that the general picture and the conclusions
do not depend on either hypothesis.
Without prior on ages, the usable sample is reduced by about
40\%  since one has to remove all non-physical ages,
but the age--metallicity relation remains, by and large, the same.
To illustrate this in Appendix \ref{noprior}, we compare the age--metallicity relation
using the ages derived with and without prior.
When ages are determined without any prior, the subset of stars with ages larger than the age
of the Universe, the non-physical subset, have a fainter lower magnitude 
than the complementary subset: brightest stars $G=13.8$ 
while they are $G=12.8$ for the complementary subset. The non-physical subset also
has larger absolute magnitudes, 2.6 for the brightest stars compared to 1.3 for the physical set.
For a given colour and metallicity, the older the age, the larger (i.e. fainter) 
the absolute magnitudes of 
the stars in the SG branch stage. Therefore, this behaviour of SPInS with and without
prior on age is understandable.

To be more quantitative in the impact of the choice of $\alpha$-enhanced
evolutionary tracks, we compared the derived ages with those obtained
using solar-scaled evolutionary tracks. The two correlate
extremely well: a linear fit provides a correlation
coefficient of 0.997, a slope very close to one
(0.983) and an offset of --0.02 Ga, in the sense that ages
derived from solar-scaled tracks are 0.02 Ga younger than those derived
from $\alpha$-enhanced tracks. The RMS around
this linear fit is 0.24 Ga. The fact that both the offset and the RMS are much smaller
than the median error in ages (2.35\,Ga) 
convinced us that the choice of the $\alpha$-enhanced
tracks does not introduce any bias.
 
We must bear in mind that the ages derived from the theoretical tracks
depend on the physics underlying the computations \citep[see e.g.][]{Lebreton2014}. By imposing
a prior, we force the derived ages to span the cosmological range
of ages.

\subsection{Test with NGC 6397}
\label{sec:ngc6397}

\begin{figure}
    \centering
    \resizebox{7.5cm}{!}{\includegraphics{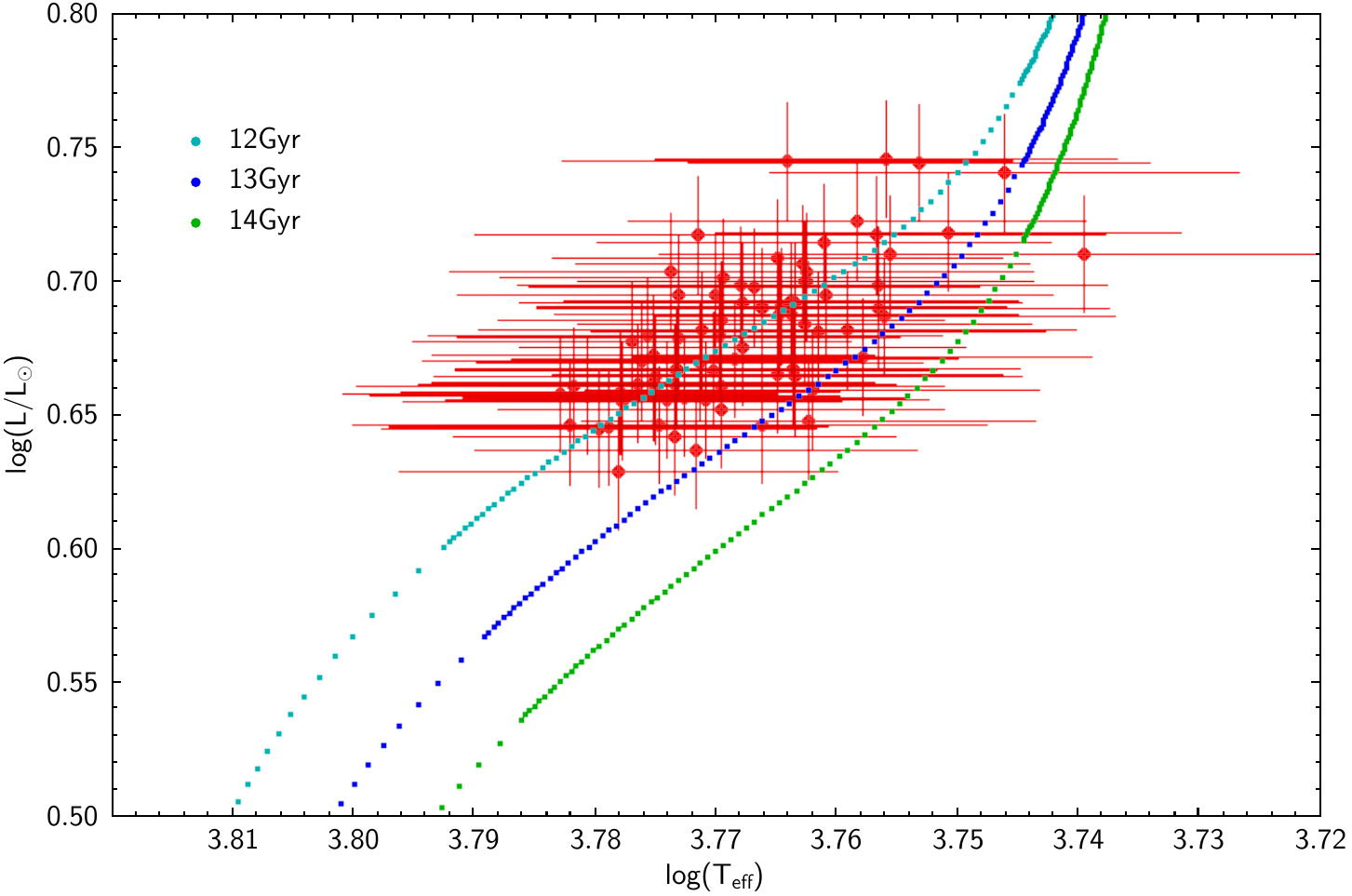}}
    \caption{Hertzsprung-Russell diagram for the SG stars of NGC\,6397 that  we used to estimate its age.
    Three BASTI isochrones of 12, 13, and 14 Ga and metallicity --2.0 are shown for reference.}
    \label{hr_ngc6397}
\end{figure}

To test SPInS in this configuration, we determined the age of
the globular cluster NGC\,6397 from
SGs in the sample of \citet{JGH}.
We cross-matched these stars with Gaia, assuming a reddening of $E(B-V)=0.186$
\citep{gratton2003}, an extinction in the $G$ magnitude of 0.79705, derived
from the KOALA grid of ATLAS 9 models \citep{koala},  
and 1.289445 in the $G_{BP}-G_{RP}$ colour as used by \citet{toposVI}.
Next we used the sub-sample of stars from \citet{toposVI}
to derive an empirical calibration between $(G_{BP}-G_{RP})_0$
and $(g-z)_0$. The best fit 
 provides $(g-z)_0 = -0.6193062+1.4829282 (G_{BP}-G_{RP})_0$.
We then used the same calibration as in \citet{toposVI} to estimate the effective
temperatures. 
We used equation \ref{l2} to estimate the luminosity of the star, assuming 
a bolometric correction for the $G$ magnitude of $-0.02352$, estimated
from the synthetic colours of the KOALA grid at metallicity --2.0, and a parallax
of $0.415\pm 0.01$\,mas \citep{VB21}. The luminosity error
was estimated from  equation \ref{l_err}. For the error on log(\teff)
we assumed an error of 250\,K on \teff.
The resulting Hertzsprung-Russell (HR) diagram is shown in Fig.\,\ref{hr_ngc6397}.
We ran SPInS on these 81 stars with the same configuration
as for our sample of field stars and obtained a mean age of 
12.3\,Ga
with a standard deviation of 
0.2 Ga
over the 81 stars. 
Since SPInS relies on a Markov chain Monte Carlo (MCMC), its output consists
of a distribution of possible ages, which implies that
for each age determination there is an associated error.
The mean error  for our 81 stars is 
0.9 Ga. 
This can be compared with the result
of \citet{C2024}, who using SPInS found a mean age of 13.292\,Ga from 83
stars in this cluster. As discussed in section \ref{comp_c24},
there are several differences between our setup of SPInS and that
used by \citet{C2024}; however, the most relevant is the prior on age.
To check this, we ran SPInS a second time without any prior on age for our sample.
In this case, the mean age was 
13.9\, Ga 
with a dispersion of 
0.5\, Ga;
the mean error was 2\,Ga, confirming that without the age prior
the errors are inflated.
Moreover, this result is closer to that of \citet{C2024}, confirming that 
the most relevant difference in our SPInS setup lies in the prior on age.

It should be noted that all of these age estimates are compatible with each other
within their errors. They are also compatible, within errors, 
with the ages found in the literature for NGC\,6397 that are
assembled in Table\,\ref{age_ngc6397}.

\begin{table}[]
    \centering
    \caption{Age estimates for NGC\,6397.}
    \renewcommand{\arraystretch}{1.5}
    \begin{tabular}{ll}
    \hline
    Age & Ref.\\
    Ga & \\
    \hline
    $13.9\phantom{00}\pm 1.1\phantom{00}$     &\citet{gratton2003}  \\
     $12.6\phantom{00}\pm 0.7\phantom{00}$    &\citet{correnti2018} \\ 
     $13.0\phantom{00}\pm 0.25\phantom{0}$   & \citet{VdB2013} \\
     $ 12.8\phantom{00}^{+0.50\phantom{0}}_{-0.75\phantom{0}}$ & \citet{torres2015}\\
     $13.292^{+0.734}_{-0.943}$ & \citet{C2024}\\
     $14.17^{+0.51}_{-0.35}$ & \citet{Massari2026}\\
     $12.3\phantom{00} \pm 0.2\phantom{00}$ & This paper, prior on ages\\
     $13.9\phantom{00} \pm 0.5\phantom{00}$ & This paper, no prior on ages\\
     \hline
    \end{tabular}
    \label{age_ngc6397}
\end{table}

\subsection{Comparison with Casamiquela et al.}
\label{comp_c24}

\citet{C2024} derived ages using SPInS for a large
sample of stars with metallicities from LAMOST \citep{Cui2012}.
Since the LAMOST low-resolution data is similar to the
SDSS \citep{York2000} spectra we used, it is reasonable to compare our results.
For this reason, we only compare this with the  Low-Resolution Spectroscopic Survey (LRS) sample of \citet{C2024}.
The main differences between our setup for SPInS and that
of \citet{C2024} are: (i) we use $\alpha$-enhanced stellar models with a constant $[\alpha/\mathrm{Fe}]=+0.4$, 
while \citet{C2024} use solar-scaled  stellar models and scale the metallicity
using the relation of \citet{salaris}; (ii) we use a prior on ages, while they use no prior;
(iii) we use log(\teff), $\log(L)$, and metallicity as input to SPInS, while
\citet{C2024} use $G_{abs}$, $G_{BP}-G_{RP}$, and metallicity.

\begin{figure}
    \centering
    \resizebox{7.5cm}{!}{\includegraphics{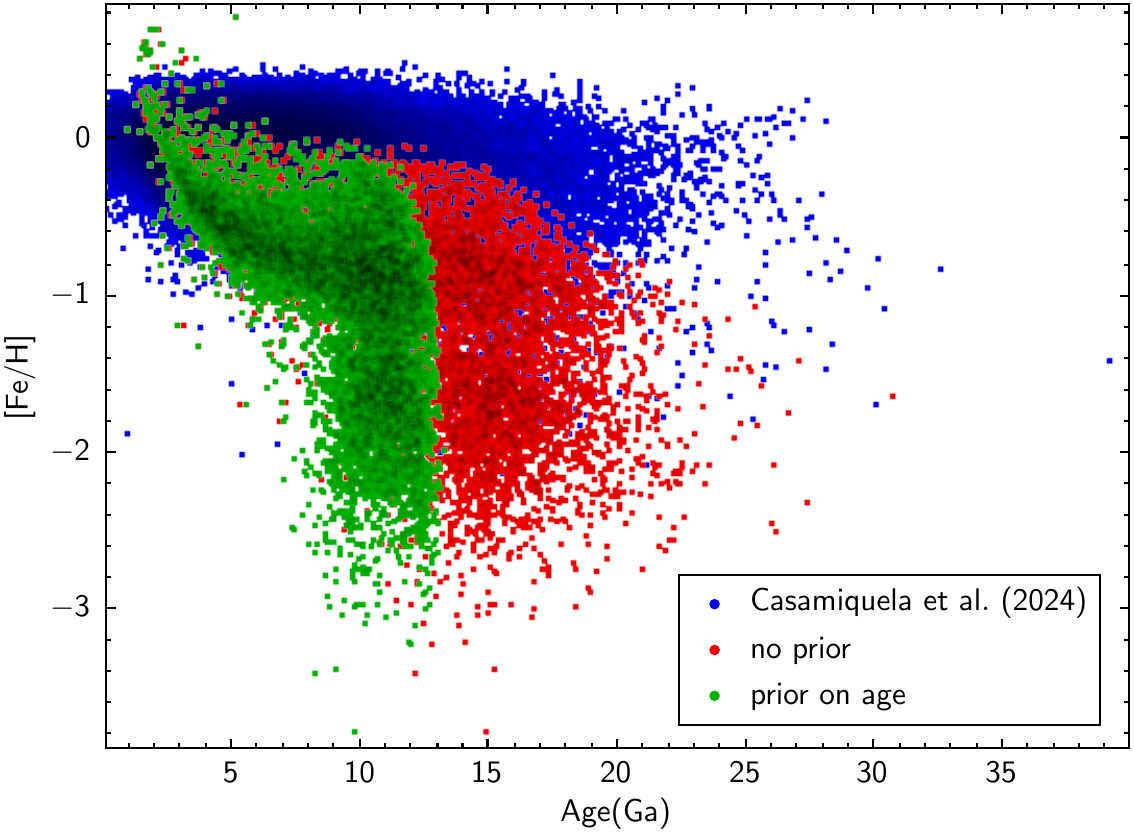}}
    \caption{AMR from our sample assuming a prior on age (green dots),
    our sample without any prior (red dots), and the sample of \citet{C2024} (blue dots). Note that there
    are red dots and blue dots below the green dots. 
}
    \label{amr_us_c24}
\end{figure}

In Fig.\,\ref{amr_us_c24} we show the comparison of the AMR
of our sample, with and without prior on age, and that of \citet{C2024}.
The effect of the prior on the ages is very obvious; however, note that for ages
smaller than the age of the Universe the morphology of the
AMR with or without prior is very similar. It is clear that our ages without prior are, by and large, consistent with those
of \citet{C2024}. The two samples differ in their metallicity distributions.

\begin{figure}
    \centering
    \resizebox{7.5cm}{!}{\includegraphics{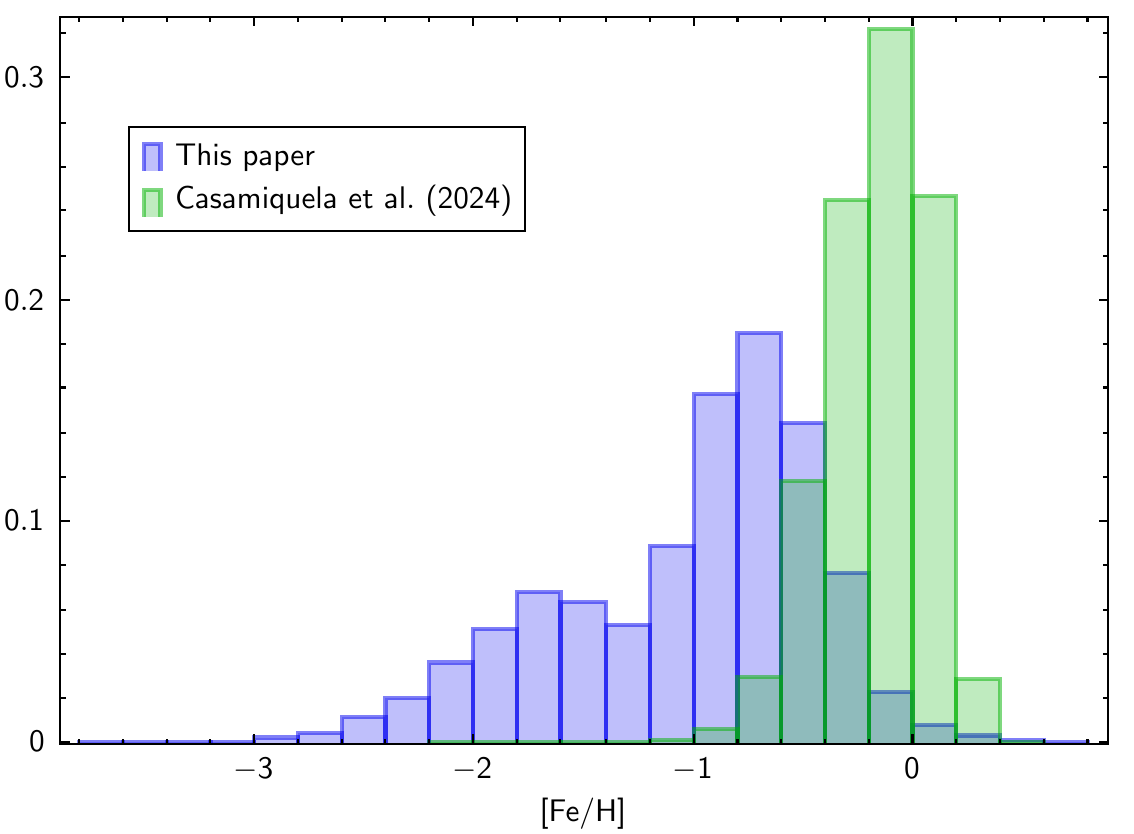}}
    \caption{Metallicity histogram of our sample (blue), compared with the LRS sample of 
    \citet{C2024} (green). The two histograms have been normalised by the number of stars
    in each sample.} 
    \label{histomet_us_c24}
\end{figure}

This is illustrated in Fig.\,\ref{histomet_us_c24}. As discussed in \citet{toposVI}, while
the SDSS sample is highly biased towards metal-poor stars, here it appears that
the LRS sample of \citet{C2024} is heavily biased towards solar metallicity
stars. Since \citet{C2024} did not introduce any metallicity cut in their sample,
this bias must arise from the LAMOST selection function.
Although we do not have kinematical data for the \citet{C2024} sample that 
allows us to classify the stars of \citet{C2024} in a way similar to what we did
for our sample, we believe that it is very likely that the sample of \citet{C2024}
is dominated by disc stars, hence their metallicity distribution.

\section{The age--metallicity relation and the assembly of the halo}

We first classified the stars dynamically.
The selections are the same as discussed in \citet{toposVI}.
Following \citet{bensby2014} we
define the stars that belong to the thin and thick disc as well as the stars that
are intermediate between the thin and thick disc.
The Toomre diagram is shown in Fig.\,\ref{toomre}.
{  As a sanity check we show in Fig.\,\ref{toomre_all} the Toomre diagram for
the whole sample of SGs, without any cut in luminosity. It can be appreciated
that, although slightly under-represented in our selection, the halo is 
well present. Therefore, we believe our selection does not miss any of the prominent components of the halo.}
We define {  the Gaia-Sausage-Enceladus (GSE) \citep{belokurov2018,haywood2018,helmi} and 
Sequoia accreted components \citep{barba2019,koppelman2019,myeong2019,villanova2019} as in \citet{feuillet2021}.
The result of the selections is shown in Fig.\,\ref{elz}.

\begin{figure}
    \centering
    \resizebox{7.5cm}{!}{\includegraphics{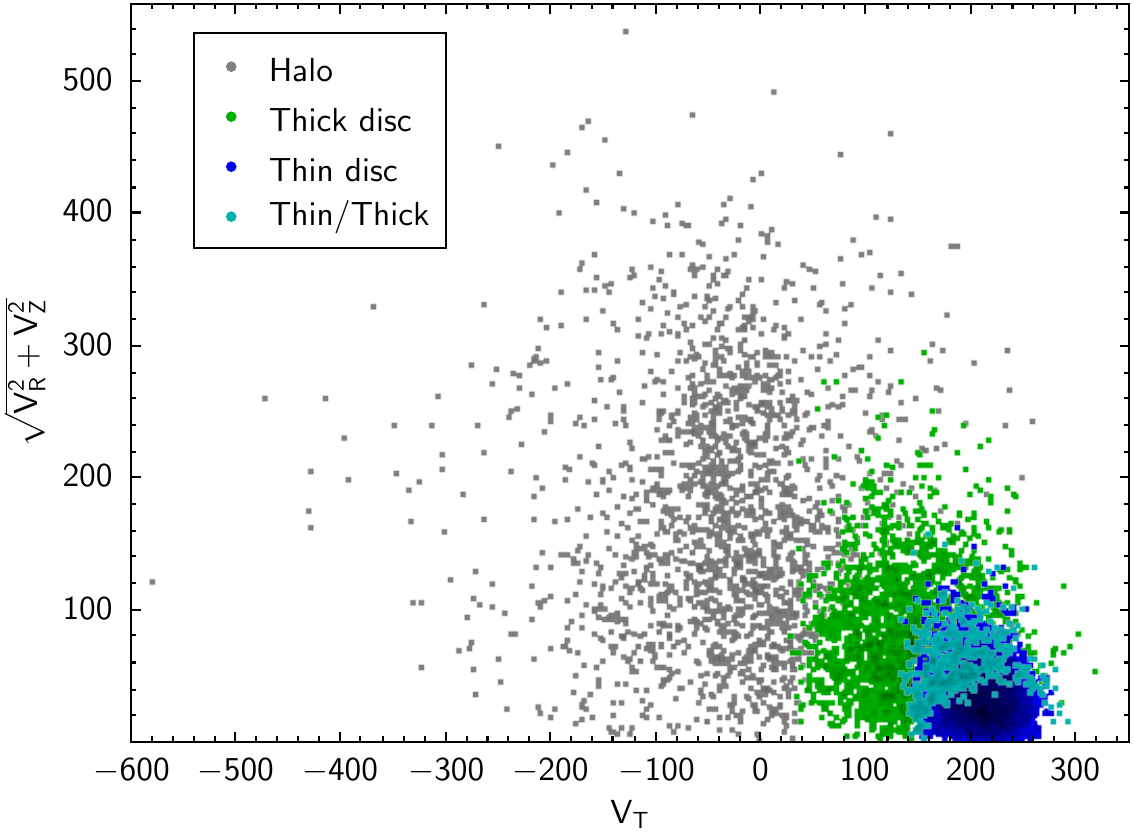}}
    \caption{Toomre diagram that we used to select thin disc and thick disc stars.}
    \label{toomre}
\end{figure}

\begin{figure}
    \centering
    \resizebox{7.5cm}{!}{\includegraphics{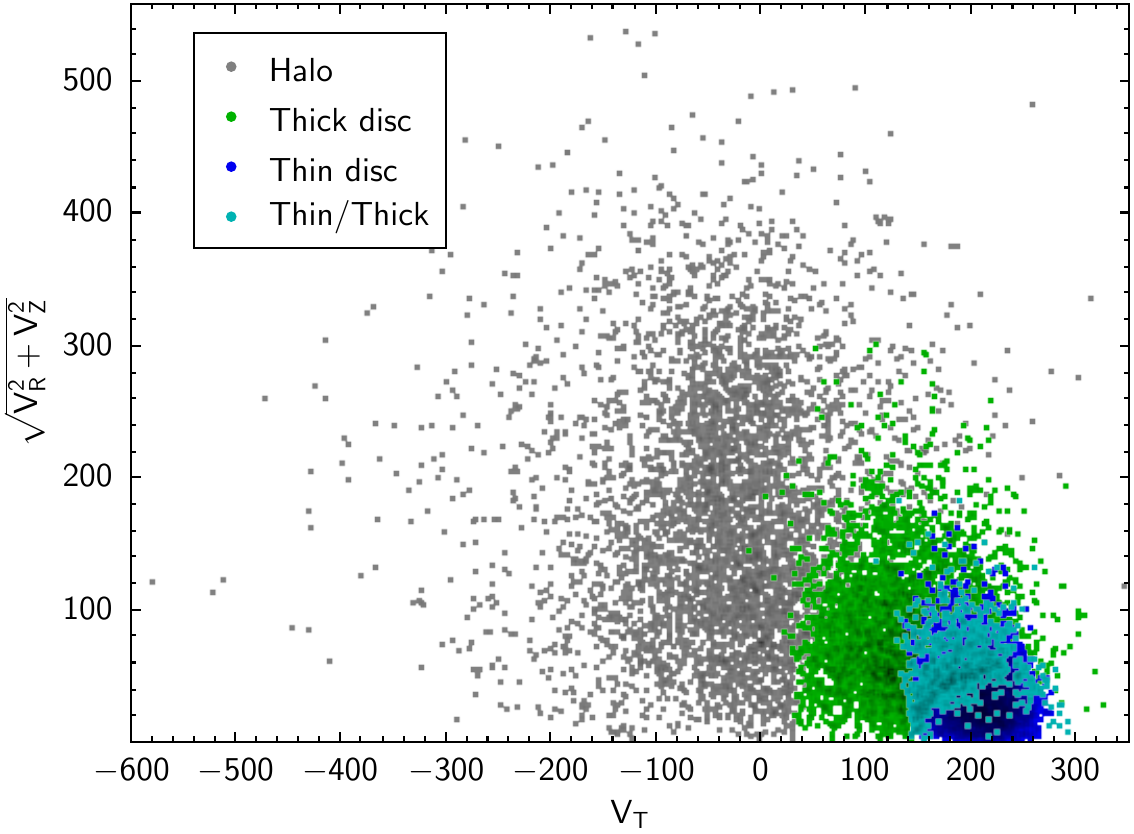}}
    \caption{Same as Fig.\ref{toomre} but for the whole sample
    of SGs, without any cut on the error in  luminosity.}
    \label{toomre_all}
\end{figure}

\begin{figure}
    \centering
    \resizebox{7.5cm}{!}{\includegraphics{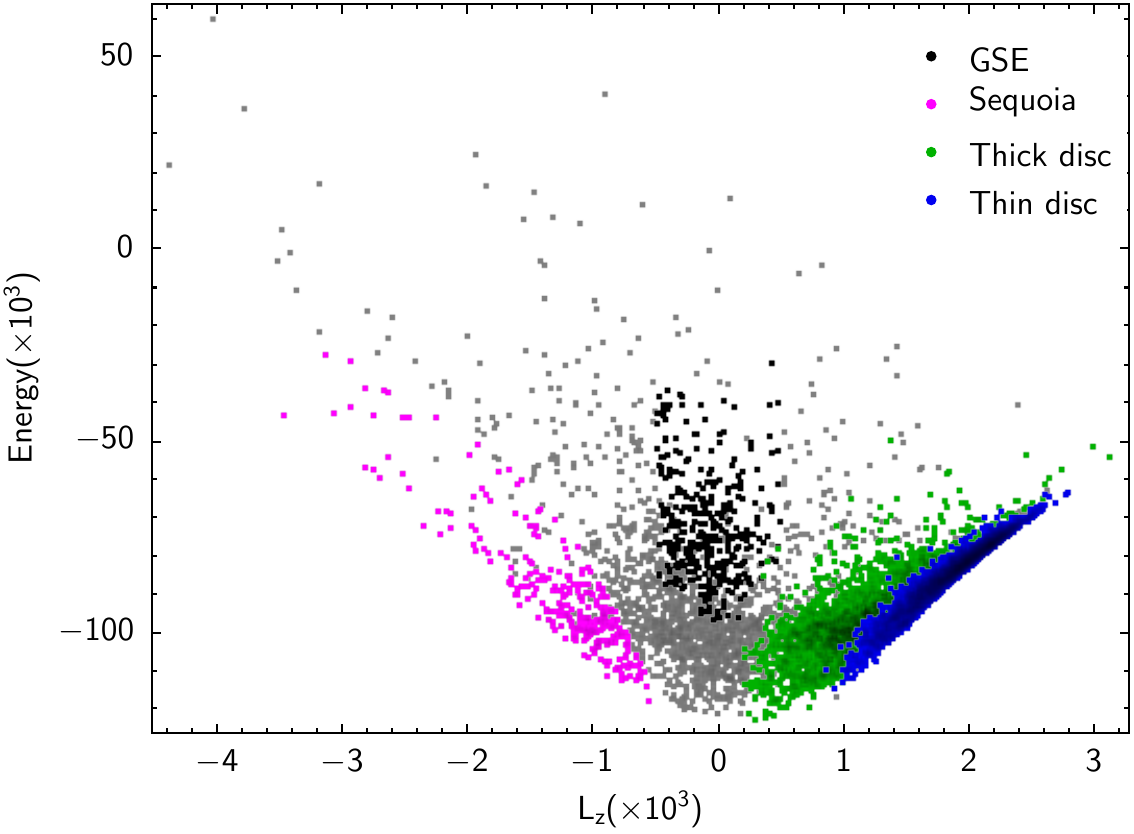}}
    \caption{Dynamically selected populations in the angular momentum--energy plane. The grey
    points are all the stars that cannot be classified as disc (thick or thin), GSE, or Sequoia.
    The grey points plus Sequoia, plus GSE, is what can generically be called halo.
 {  The units for energy are    
 $\rm 10^3\times km^2\,s^2$ (actually it is a specific energy, or energy per unit mass)
 and
 $\rm 10^3\times kpc\, km\,s^{-1}$ for the angular momentum. 
 }
 }
    \label{elz}
\end{figure}

\begin{figure}
    \centering
    \resizebox{7.5cm}{!}{\includegraphics{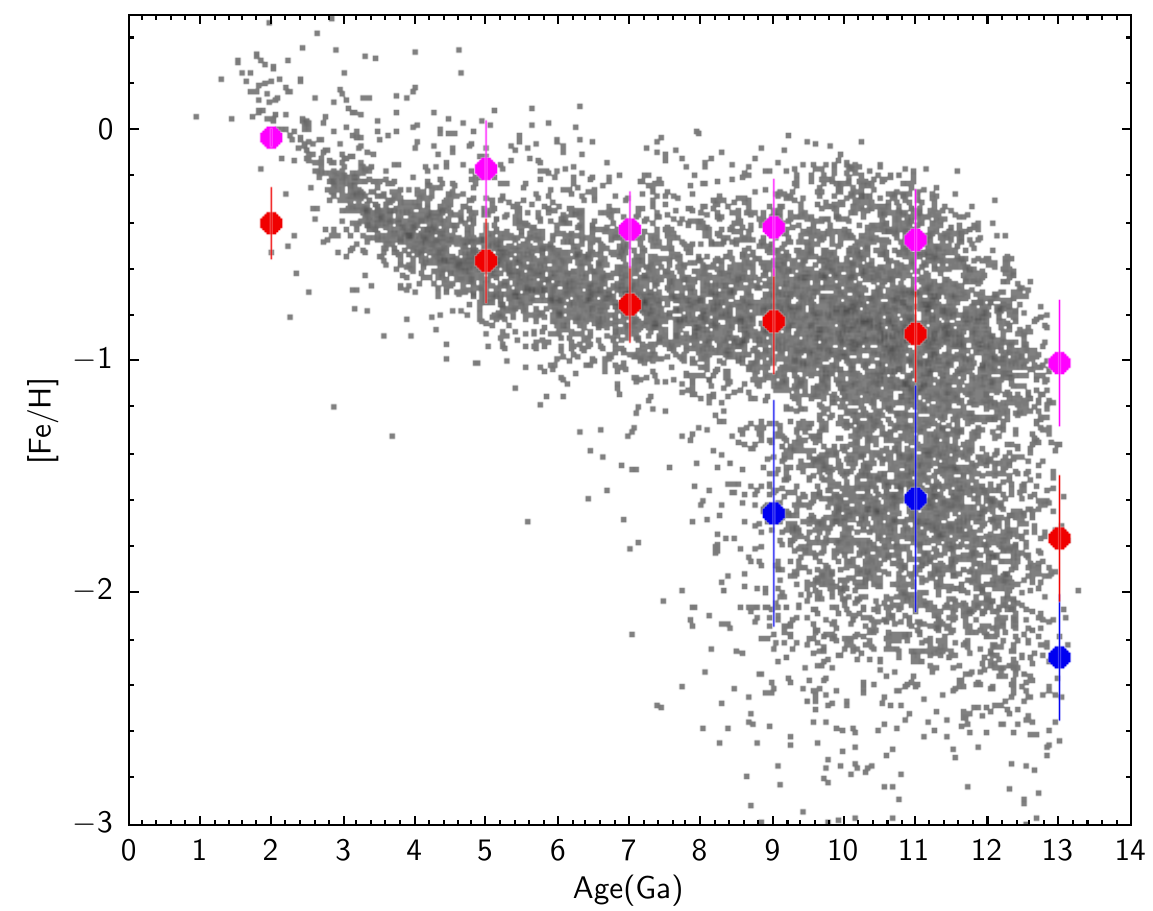}}
    \caption{AMR as defined from our dataset.
    The data are divided into six age bins, and for each bin we show
    the points corresponding to the centre of the bin and to the metallicity of the peaks.
    The vertical bars correspond to the FWHM of a Gaussian to each peak. The blue dots
    correspond to the low-AMR, the red dots to the mid-AMR, and the purple points to the
    high-AMR.}
    \label{agemet_peaks}
\end{figure}

\begin{figure}
    \centering
    \resizebox{7.5cm}{!}{\includegraphics{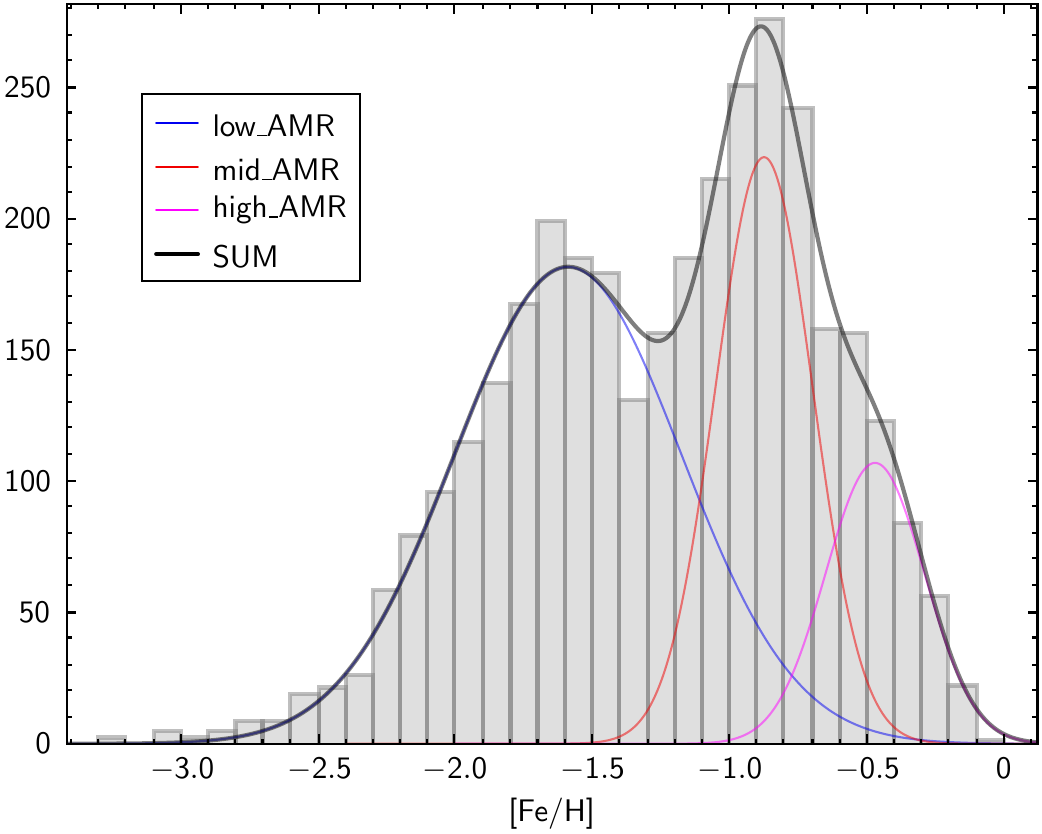}}
    \caption{Histogram of the metallicities in the  age bin 10 to 12 Ga. The fitted Gaussians
    are shown both individually (blue, red, magenta lines) and summed (black line).}
    \label{a10_12histo}
\end{figure}

\subsection{Different age--metallicity relations}\label{sec:diff_amr}
We can now look at the AMR shown in Fig.\,\ref{agemet_peaks}.
  At this stage, we may wonder if our selection has in some way biased our diagram.
At any given metallicity, the younger populations are found among turn-off and SG 
stars. Since our selection favours these evolutionary stages for metal-poor stars
bias would manifest as an excess of young stars at low metallicity. Since this
is not the case, we conclude that our selection does not introduce any significant
bias in this diagram. The probable reason is that the stars in our selection that are not
SGs are a minority.
To help the interpretation of the diagram we divided the sample
into six age bins, and for each bin we noted the metallicity of obvious peaks
in the metallicity histogram and their FWHM, both estimated by a Gaussian fit. 
The histograms show three, partially overlapping, peaks in the three lowest metallicity bins  
and only two peaks for the remaining metallicity bins. 
As an example, in Fig.\,\ref{a10_12histo} we show the histogram of the metallicities
in the age bin
10\,Ga to 12\,Ga.
We are well aware that these multiple peaks appear only due 
to the bias of the SDSS spectra
boosting the number of stars of low metallicity
\citep[see figure 10 of][]{toposVI}.
However, here we are not trying to derive a metallicity distribution function,
and we do not make any inference from the relative number of stars in any peak.
We are trying to infer the AMR, and it is fairly obvious that
our sample suggests at least three AMRs. 
For ease of discussion, in the following we refer to 
the three as low-AMR, mid-AMR, and high-AMR, respectively.

To gain further insight let us look at the same plot, but this time including only the
stars classified as GSE (Fig.\,\ref{agemet_gse})
and thick disc   (Fig.\,\ref{agemet_td}).
In the first case, it is obvious that the GSE stars follow the low-AMR.
The thick disc instead comprises both stars that follow the low-AMR 
and stars that follow the mid-AMR. 
The thin disc stars follow in part the high-AMR
and in part the mid-AMR (see Fig.\,\ref{agemet_thin}).
If we look at the {  age--metallicity }plot with stars classified as Sequoia, they
also follow by and large the low-AMR, and perhaps are defining an
AMR at even lower metallicity (Fig.\,\ref{agemet_seq}).
The stars classified as halo, which also include all the GSE and Sequoia stars, 
follow the low-AMR, with a minority of stars that remain in the mid-AMR (Fig.\,\ref{agemet_halo}).

\begin{table}[]
    \centering
    \caption{Percentage of stars classified as thin disc, thick disc, and halo along with each of the three AMRs.}
\begin{tabular}{rrrr}
\hline
  \multicolumn{1}{c}{AMR} &
  \multicolumn{1}{c}{Thin Disc (\%)} &
  \multicolumn{1}{c}{Thick Disc (\%)} &
  \multicolumn{1}{c}{Halo (\%)} \\
\hline
low-AMR & 0.16\% & 21.16\% & 78.68\%\\
 mid-AMR & 3.37\% & 26.81\% & 69.83\%\\
 high-AMR & 68.57\% & 26.78\% & 4.65\%\\
\hline\end{tabular}
    \label{tab:percentage}
\end{table}

\begin{figure}
    \centering
    \resizebox{7.5cm}{!}{\includegraphics{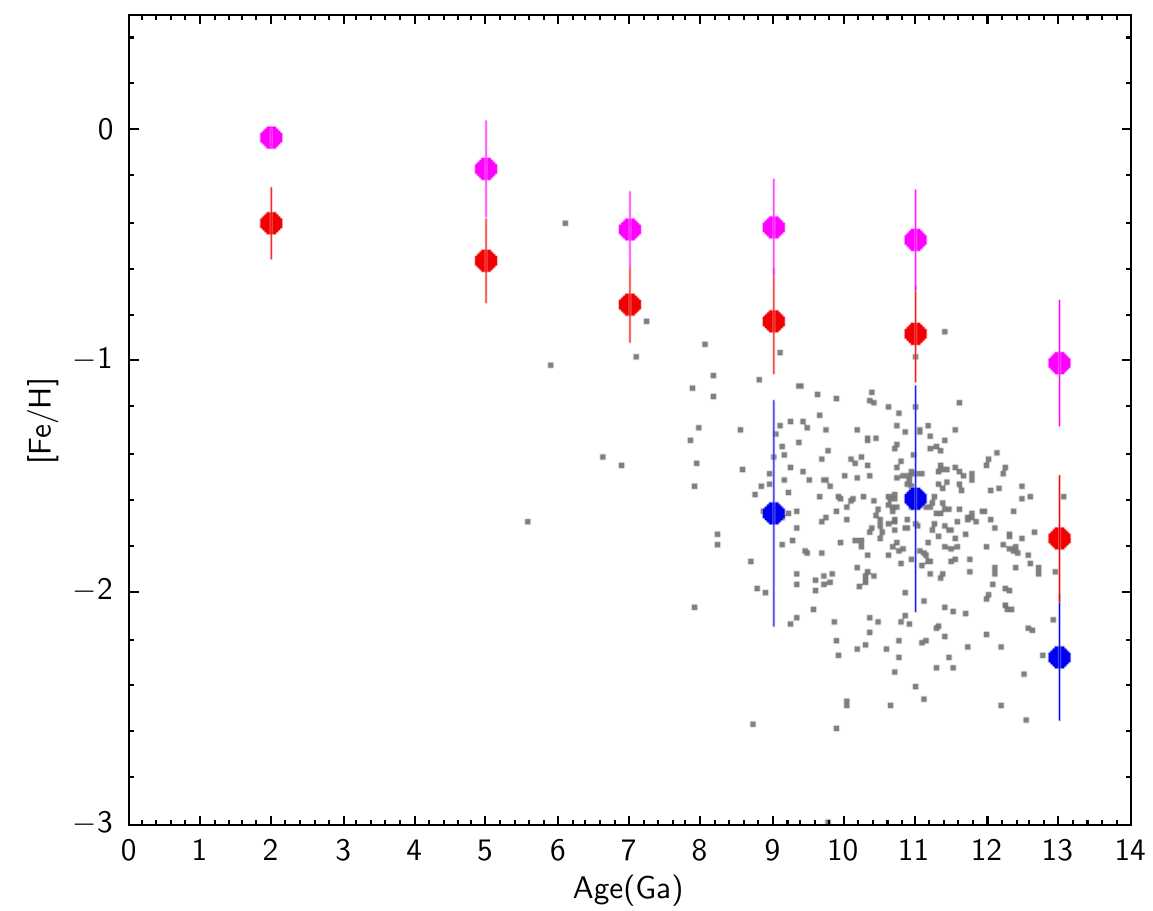}}
    \caption{As Fig.\,\ref{agemet_peaks}, but now only the stars
    classified as GSE are shown. 
    The meaning of the colours of the dots is the same as
    in Fig.\,\ref{agemet_peaks}.
     }
    \label{agemet_gse}
\end{figure}

\begin{figure}
    \centering
    \resizebox{7.5cm}{!}{\includegraphics{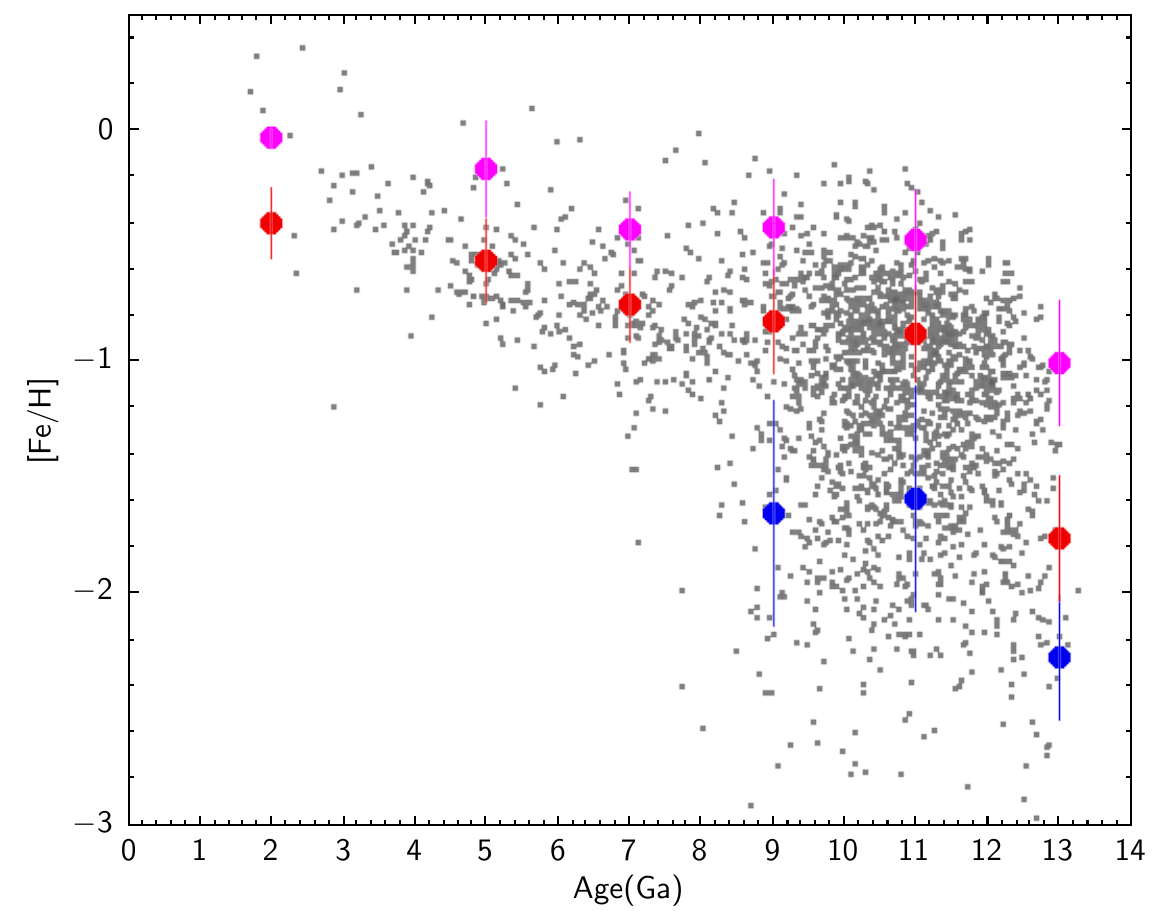}}
    \caption{As Fig.\,\ref{agemet_peaks}, but now only the stars
    classified as thick disc are shown.
    The meaning of the colours of the dots is the same as
    in Fig.\,\ref{agemet_peaks}.
     }
    \label{agemet_td}
\end{figure}

\begin{figure}
    \centering
    \resizebox{7.5cm}{!}{\includegraphics{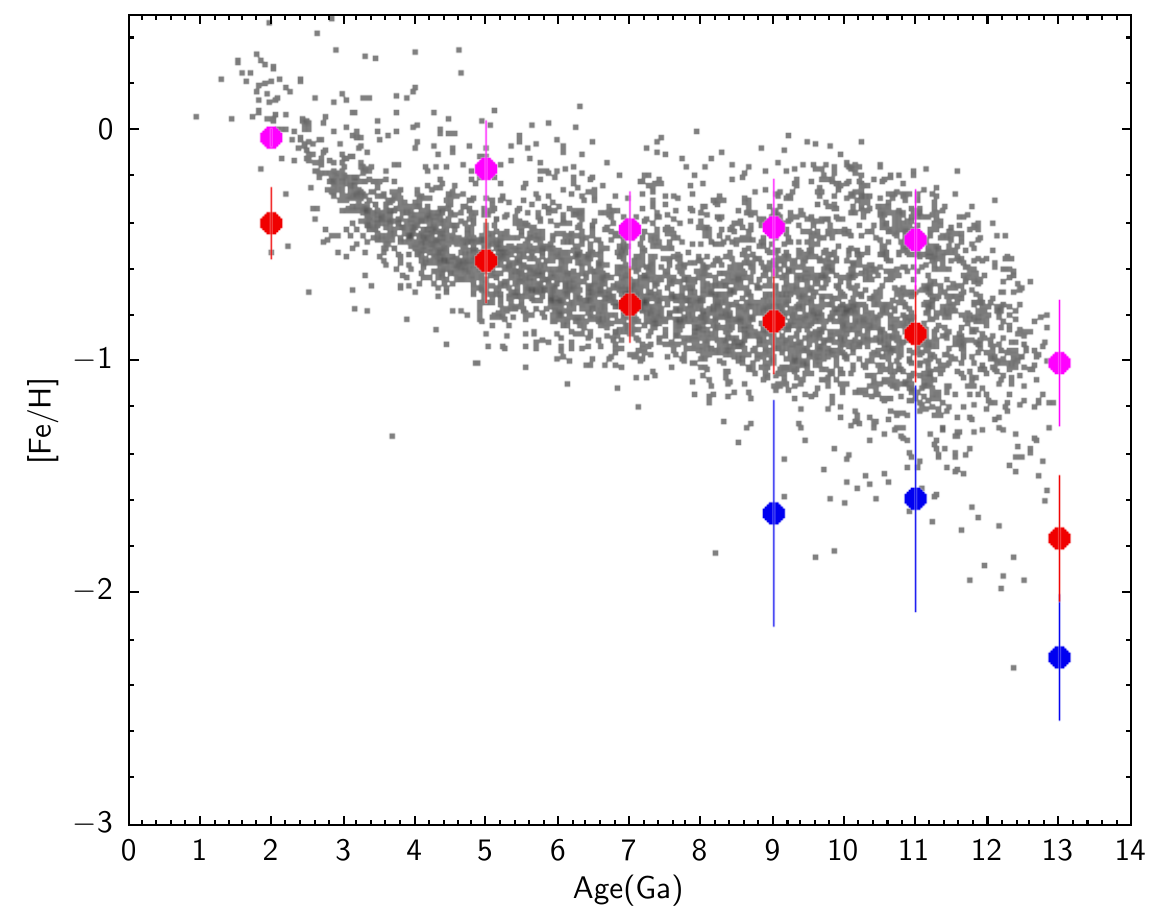}}
    \caption{As Fig.\,\ref{agemet_peaks}, but now only the stars
    classified as thin disc are shown.
    The meaning of the colours of the dots is the same as
    in Fig.\,\ref{agemet_peaks}.
     }
    \label{agemet_thin}
\end{figure}

\begin{figure}
    \centering
    \resizebox{7.5cm}{!}{\includegraphics{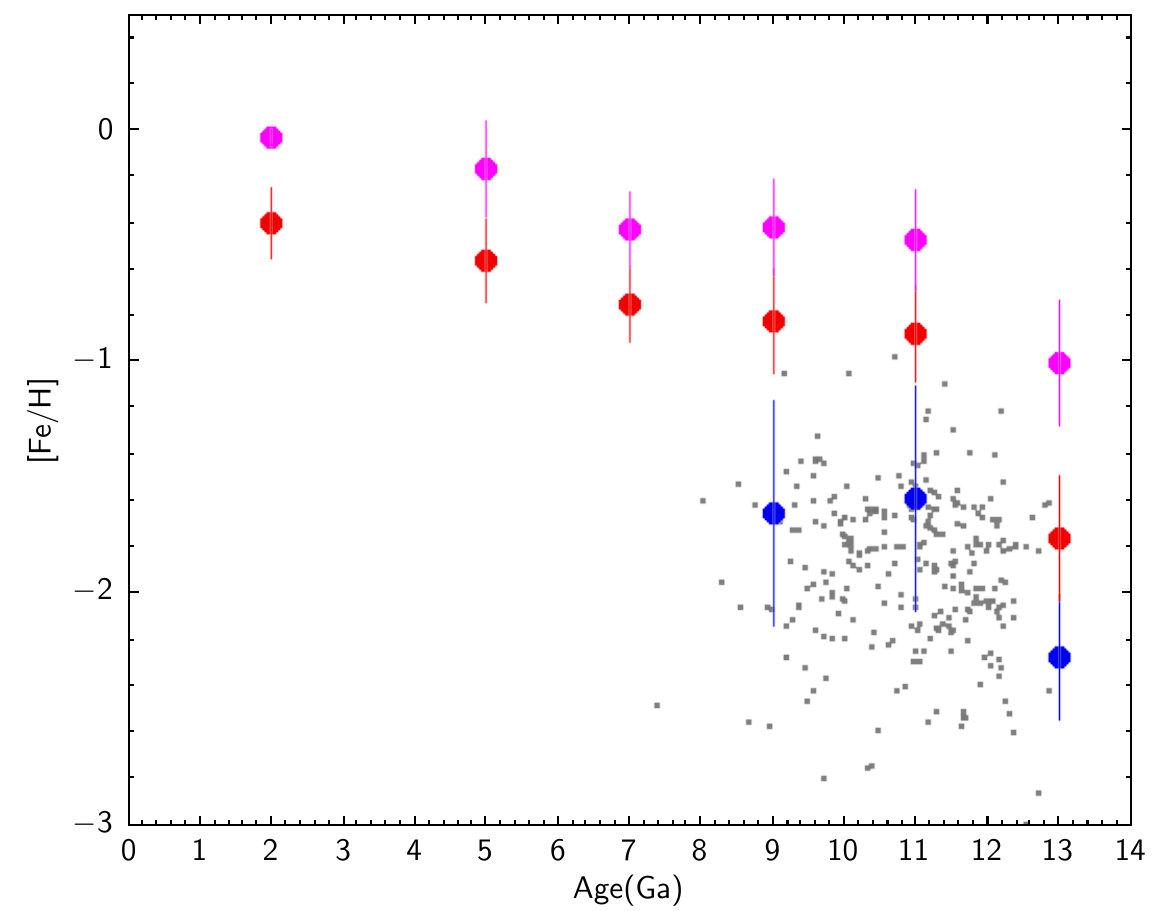}}
    \caption{As Fig.\,\ref{agemet_peaks}, but now only the stars
    classified as Sequoia are shown.
    The meaning of the colours of the dots is the same as
    in Fig.\,\ref{agemet_peaks}.
     }
    \label{agemet_seq}
\end{figure}

\begin{figure}
    \centering
    \resizebox{7.5cm}{!}{\includegraphics{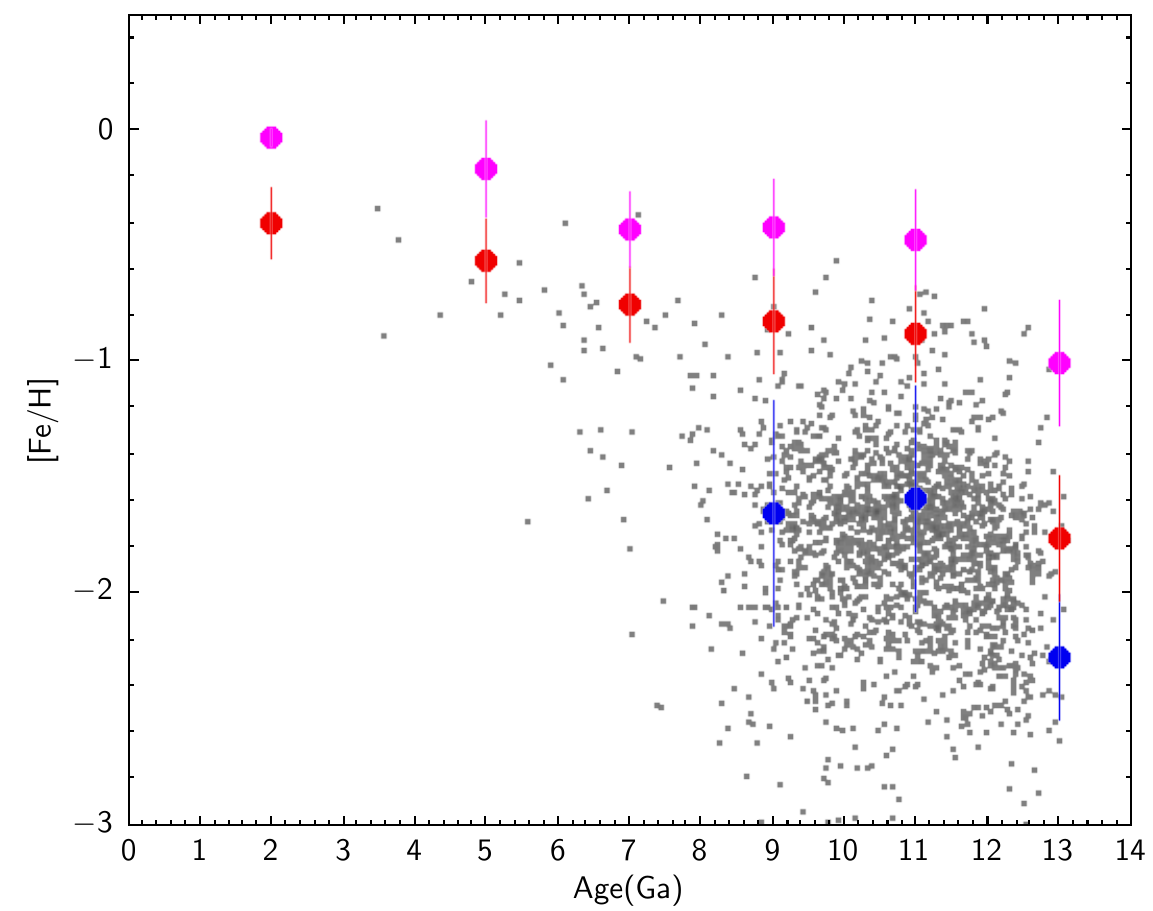}}
    \caption{As Fig.\,\ref{agemet_peaks}, but now only the stars
    classified as halo are shown.
    The meaning of the colours of the dots is the same as
    in Fig.\,\ref{agemet_peaks}.
     }
    \label{agemet_halo}
\end{figure}

We assigned to each star the probability
of belonging to each of the three AMRs, using the Gaussian fits in each age bin, 
and then assigned it to the AMR to which it had the 
highest probability of belonging.
In Table \ref{tab:percentage} we provide the percentages of 
the components in each AMR. It is clear that the high-AMR is
dominated by the thin disc and the low-AMR is dominated by the halo.
The mid-AMR is also dominated by the halo, but slightly less than the low-AMR
and with a slightly larger fraction of disc stars, mainly thick but also a few thin. 
The thick disc stars are distributed
almost equally among the  three AMRs, for each of which they constitute in the order of one-quarter
of the stars.

\citet{belokurov2018} estimate that the GSE accretion event occurred between 8 and 11 Ga ago,
while our GSE stars span a larger range extending to more than 13\,Ga. These two facts are not
contradictory: in fact, it is expected that the GSE progenitor was forming
stars well before the accretion event. On the other hand, the fact that we do not find stars younger
than 8\,Ga is consistent with the accretion time estimated by \citet{belokurov2018}, since after the
accretion GSE must have lost its gas and stopped forming stars.

Our AMR is morphologically similar to that provided
by \citet[][see their Figure 3, left panel]{gallart2019}, although there is
a clear shift in age and metallicity between the two. For metallicities
the shift can be easily ascribed to the different techniques used: in our case the analysis
of spectra, in the case of \citet{gallart2019} the analysis of the colour magnitude diagram (CMD)
and the different samples used. In the sample of \citet{gallart2019} the most metal-poor stars
are at metallicity around --1.50, whereas in our sample we go down to metallicity --3.0.
This is due to the increase of the number of metal-poor stars in the SDSS spectroscopic
sample, already discussed by \citet{toposVI}.
The shift in ages can be ascribed to two different factors. The first is methodological, as
in  \citet{gallart2019} their CMD fitting procedure
provides as output  the age--metallicity  distribution and the SFR as a function of time.
This procedure suffers from age--metallicity degeneracy, as old metal-poor stars can
have the same position in the CMD as younger stars of higher metallicity. However,
fitting the whole CMD, from below the TO to the tip of Red Giant Branch (RGB) 
minimises this degeneracy. 
In our case, we do not have a degeneracy because of the spectroscopic determination of metallicities.
The other difference is the selection process: while the sample of \citet{gallart2019}
is limited to a sphere of about 2\,kpc, centred on the Sun, our sample covers a larger volume
and is not symmetric, having more stars towards the Galactic anticentre than towards the Galactic
centre. Both of these facts could contribute to the fact that our sample has a larger number of stars younger
than 6\,Ga, which are very rare in the sample of  \citet{gallart2019}. 
Nevertheless, when we look at Figure 3 of \citet{gallart2019} it is fairly obvious that with increasing
age the spread of metallicity increases, which is what we mean by saying that the two age-metalllicity
relations are morphologically similar.

Another  striking thing in Fig.\,\ref{agemet_peaks} is that if we look at the data
for individual stars with ages
older than about 8 Ga there does not appear to be a clear correlation between age and
metallicity, while at younger ages there appears to be a clear correlation. 
It is due to the analysis of the histograms and their peaks that we were able to define
the three AMRs described above.

To assess the robustness of the multi-modality of the metallicity distributions in each age bin
we analysed the three 
oldest age bins, where multiple peaks are visible in the metallicity histogram: 8--10\,Ga, 10--12\,Ga,
and 12-14\,Ga using routines from the {\tt scikit-learn} library \citep{scikit-learn}. Each bin was subject to a bootstrapping analysis. In each bin the original data was used
to generate a new dataset of the same size, obtained by random selection in the original set, with replacement.
This means that in each bootstrap sample there are some duplicated points and some points are missing.
We computed 100 bootstrap samples for each age bin. In each of the bootstrap samples three peaks were present,
although in the 12--14\,Ga age bin the three peaks are less distinct. 
We then proceeded to an error-aware bootstrap analysis; in this case, each point was not simply picked
up from the original dataset but was modified by adding to it a Gaussian error that was randomly picked
among the available errors. Although the histogram results are smoothed by this exercise, 
the three-peak structure persists. As a further test we used a Gaussian mixture model (GMM). 
For each age bin we used both the Akaike information criterion \citep[AIC]{akaike1974}
and the Bayesian information criterion \citep[][BIC]{schwarz1978} to decide 
whether the two- or three-component model was to be preferred. In each case, the AIC and BIC
values were very close to each other, giving little discriminating power. 
Furthermore, the results were mixed: for the age bin 8-10\,Ga three components were preferred
by both criteria; for the 10-12\,Ga age bin AIC favours three components and BIC prefers two components;
finally for the bin with ages $> 12$\,Ga both criteria prefer the two-component model.
What is of essence here is that in any case the GMM analysis supports the existence of
 multi-modality in the data.
Since the GMM is of little help in deciding whether  two or three components
are preferred, we assume that the third peak is also a 
distinct component, as indicated by both bootstrapping analyses.

\subsection{A plausible scenario}

We want to present a scenario that may explain the current observations.
In the first 4\,Ga the evolution of the Milky Way saw the formation of a disc, but also 
a large number of minor mergers, like Sequoia. Each of these dwarf galaxies
had its own star formation history and metallicity distribution \citep[see e.g.][]{Salvadori2015}.
This explains the large dispersion of the metallicities observed in this age range.
Each of the merging galaxies had its AMR and the
low-AMR is  the average of all these relations. To complicate the picture,
we must also consider that the merging galaxies were gas-rich. 
Starbursts were likely triggered in the merging process
\citep[see e.g.][]{Yang}, further increasing the spread of metallicity,
extending it to higher metallicities than those found in the merging galaxy
prior to the beginning of the merging process.

The fact that the mid-AMR and high-AMR seem to be associated with the discs
suggests that they trace the growth of metallicity in these two major
structures of the Milky Way.
This also suggests that the disc was not homogeneous but that 
there were regions of the disc that were increasing their
metallicity at higher rates than others. The mid-AMR and high-AMR are an average
of the lower and higher rates of growth in metallicity.

The collision
between the Milky Way and the GSE progenitor, a major merger, occurred about 8\,Ga to 10\,Ga ago. 
This event likely triggered a starburst both in the incoming progenitor \citep[see e.g.][]{Wang2024},
and in the Milky Way, whose disc was heated and formed what is now
the thick disc. Mergers always trigger a starburst, although the associated star formation 
may be a small
fraction of the total SFR of the galaxy \citep{Wang2019}.
It is possible that the arrival of the GSE progenitor  also  had an effect on the
population of dwarf galaxies and star clusters that were powering the minor
merger episodes,
either by accelerating 
their merger with the Milky Way or ejecting them so that they became unbound. 
  
Although the present data do not directly support a causal connection
between the decrease in metallicity dispersion and the GSE merging, 
the temporal coincidence of the two facts is intriguing and suggests
speculation that they are indeed connected.
 
Merger simulations with sufficient spatial resolution 
are needed to assess the viability of this scenario.
For younger ages our sample seems to be entirely constituted by disc stars.

The fact that a few of the stars of the thick disc, at low metallicities,  follow  the low-AMR 
suggests that some of the  stars accreted in the merger process ended up in disc orbits
as also suggested by \citet{mori}.
In the same way, the fact that some of the halo stars follow the high-AMR
suggests that in the collision processes some of the stars
originally in the disc ended up in halo orbits.
This is consistent with the finding of \citet{belokurov2020}
of the so-called {\em Splash} population, which is made up of metal-rich
stars (metallicity larger than --0.7) in halo orbits. Although for
\citet{belokurov2020} it was possible to isolate the metal-rich component of this population, 
because they used APOGEE as a source of metallicities, there is no physical reason
why more metal-poor stars, if present in the disc at the time of the perturbing merger,
should not also be sent into halo orbits. In our sample we highlight
the presence of such stars at any metallicity.
It is important to remember 
that our dynamical groups are defined independently of the
chemical composition.
This scenario also explains why purely kinematical, purely chemical,
or mixed selections of Galactic components define different sets of stars
as shown by \citet{franchini2020}. Last but not least, it is worth noting that part 
of the scatter in the AMR for ages older than 8~Ga 
and metallicities lower than [Fe/H]$=-1$ may be 
driven by inhomogeneous enrichment during the 
earliest stages of galaxy formation \citep[e.g.,][]{Vanni2023, Rossi2024}.

\begin{figure}
    \centering
    \resizebox{7.5cm}{!}{\includegraphics{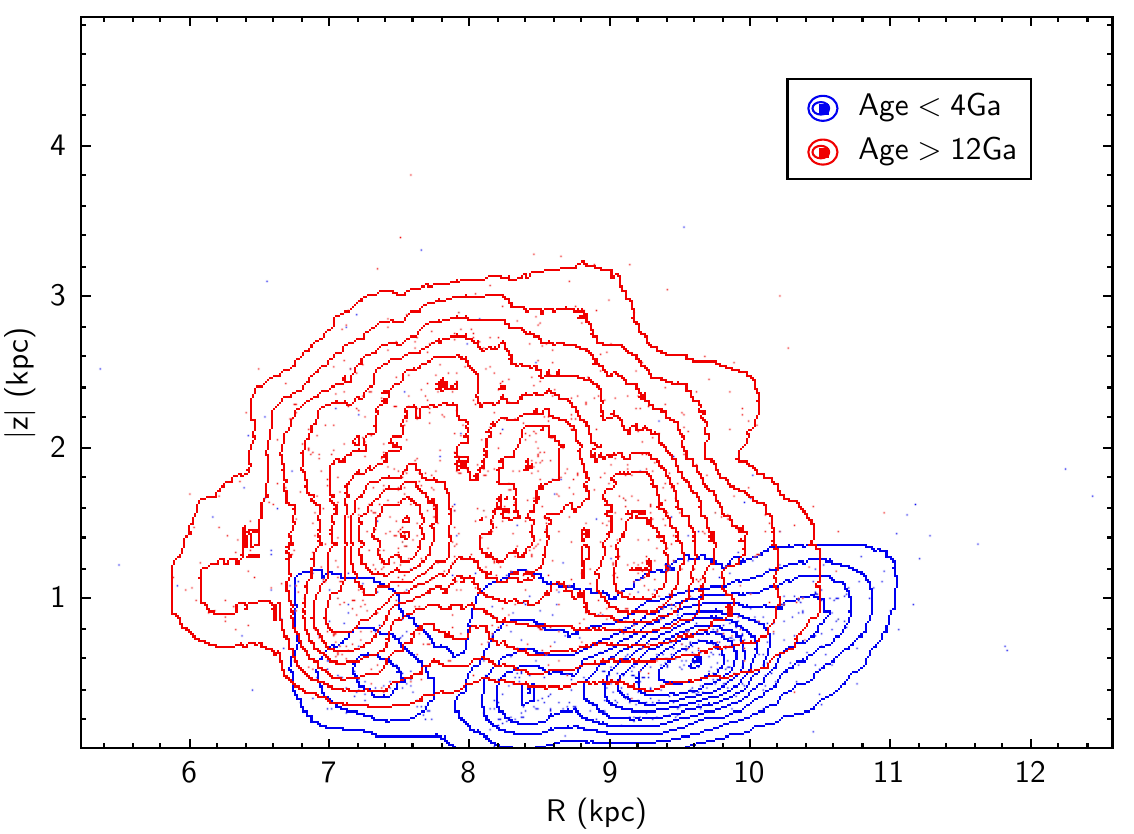}}
    \caption{Contour plot of the two extreme age bins (youngest and oldest)
    in the $R, |z|$ plane.}
    \label{map_age}
\end{figure}

In Fig.\,\ref{map_age} we show for the two extreme age bins (youngest, blue, and
oldest, red) the contour plots in the plane of the Galactocentric cylindrical radius and
the absolute value of $z$. This shows clearly that the younger stars in our sample
tend to concentrate at lower $|z|$ and preferentially at Galactocentric radii
larger than the Sun, consistent with an inside-out formation scenario for Galactic
discs \citep[e.g.][]{larson1976,baker2025}.

\subsection{Blue metal-poor stars}

\begin{figure}
    \centering
    \resizebox{7.7cm}{!}{\includegraphics{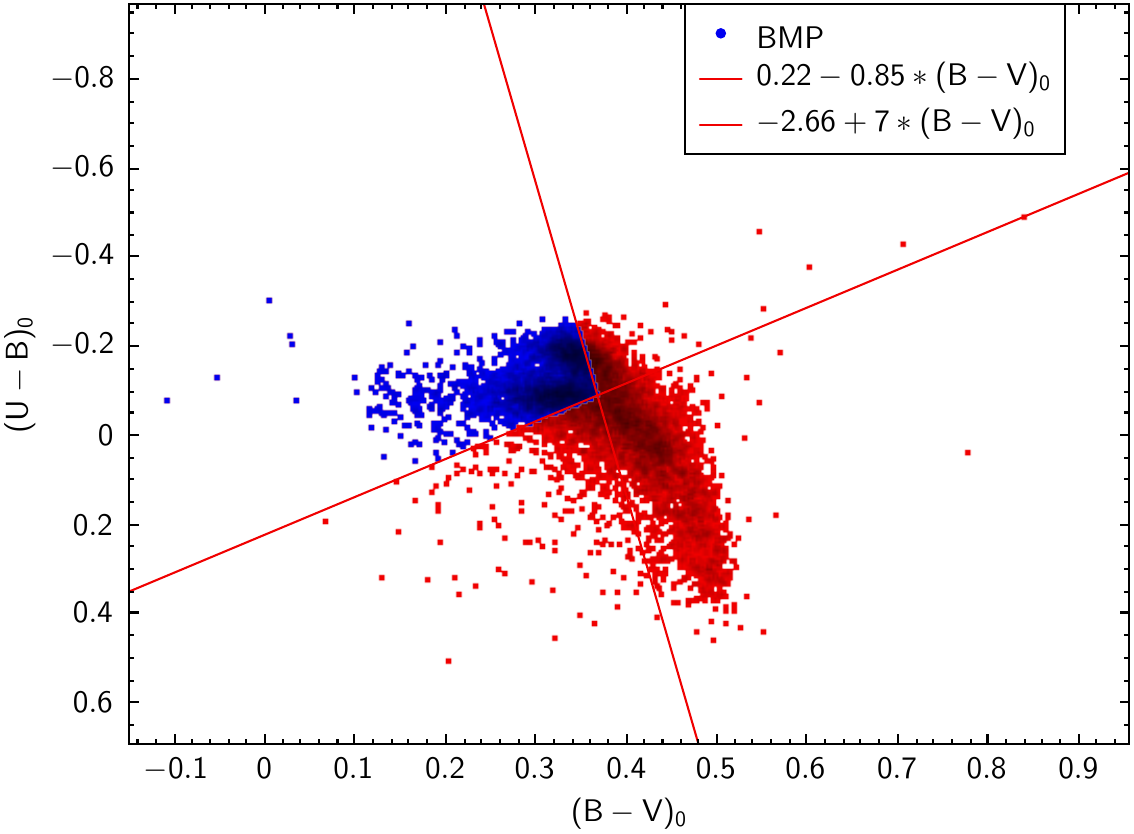}}
    \caption{Stars in our sample that classify as blue metal-poor according to the criteria of \citet{Preston1994}.}
    \label{fig:bmp}
\end{figure}

For a long time, astronomers have been intrigued by the population of 'blue metal-poor' (BMP, hereafter) stars,
which should contain a mixture of blue straggler stars (BSS hereafter) and intermediate age
metal-poor populations. 
\citet{Preston1994} wanted to select metal-poor field stars with main sequence  
gravities, thus excluding horizontal branch stars, and to do so they introduced
a photometric selection criterion 
based on Johnson $UBV$ bands. 
In our case, we transformed the 
SDSS photometry to Johnson photometry\footnote{\url{https://www.sdss4.org/dr17/algorithms/sdssUBVRITransform/}}
and corrected for reddening using the extinction in the SDSS catalogue.
The selection of BMP stars is shown in Fig.\,\ref{fig:bmp} and it consists
of 3583 stars, which is 41\% of the sample.
The first consideration is that the majority of BMP stars are 
metal-poor (MP;\footnote{We adopt the definitions of metal-poor (MP), very metal-poor (VMP), extremely metal-poor (EMP), and ultra metal-poor 
(UMP) stars from \citet{Bonifacio2025}, Table 1.}, 52\% have $-1.5 \le$ [Fe/H] $\le -0.5$).
In the second place young metal-poor stars, or BSS, defined
as BMP stars with age $\le 10\,$Ga and MP,
are only 46\% of the BMP stars. 
This fraction rises to 57\% if we also add the VMP and EMP stars 
to the sample.
These stars  have masses between 0.78~M$_\sun$\
and 1.4~M$_\sun$, and 10\% of these have masses larger than 1.0~M$_\sun$.
These characteristics are compatible with the BSS nature of these stars.
In the sample of \citet{Preston2000} the BSS were estimated to be 50\% of the sample,
which matches very well the 46\% or 57\% found in our sample.
The BMP stars  are not necessarily young or BSS. In fact, 
29\% of our BMP sample have ages older than 10\,Ga.
Finally, in our sample, there is a total lack of young metal-poor stars,
with masses above 1.6~M$_\sun$\ that can hardly be interpreted as BSS
without invoking fusions of triple systems \citep{Bonifacio2024}.
Several such stars have been found in the high speed sample of 
\citet{Bonifacio2024} and in the high radial velocity samples
of \citet{Caffau2024,Caffau2025}, and \citet{Katz2025}.
We interpret this fact as due to the rarity of this population, that
we selected a rapid evolutionary phase (SGs), 
and because our sample is small.

\section{Comparison with chemical evolution models of the Milky Way and dwarf galaxies}
\label{sec:chem}

\begin{figure}
    \centering
    \resizebox{0.4\textwidth}{!}{\includegraphics{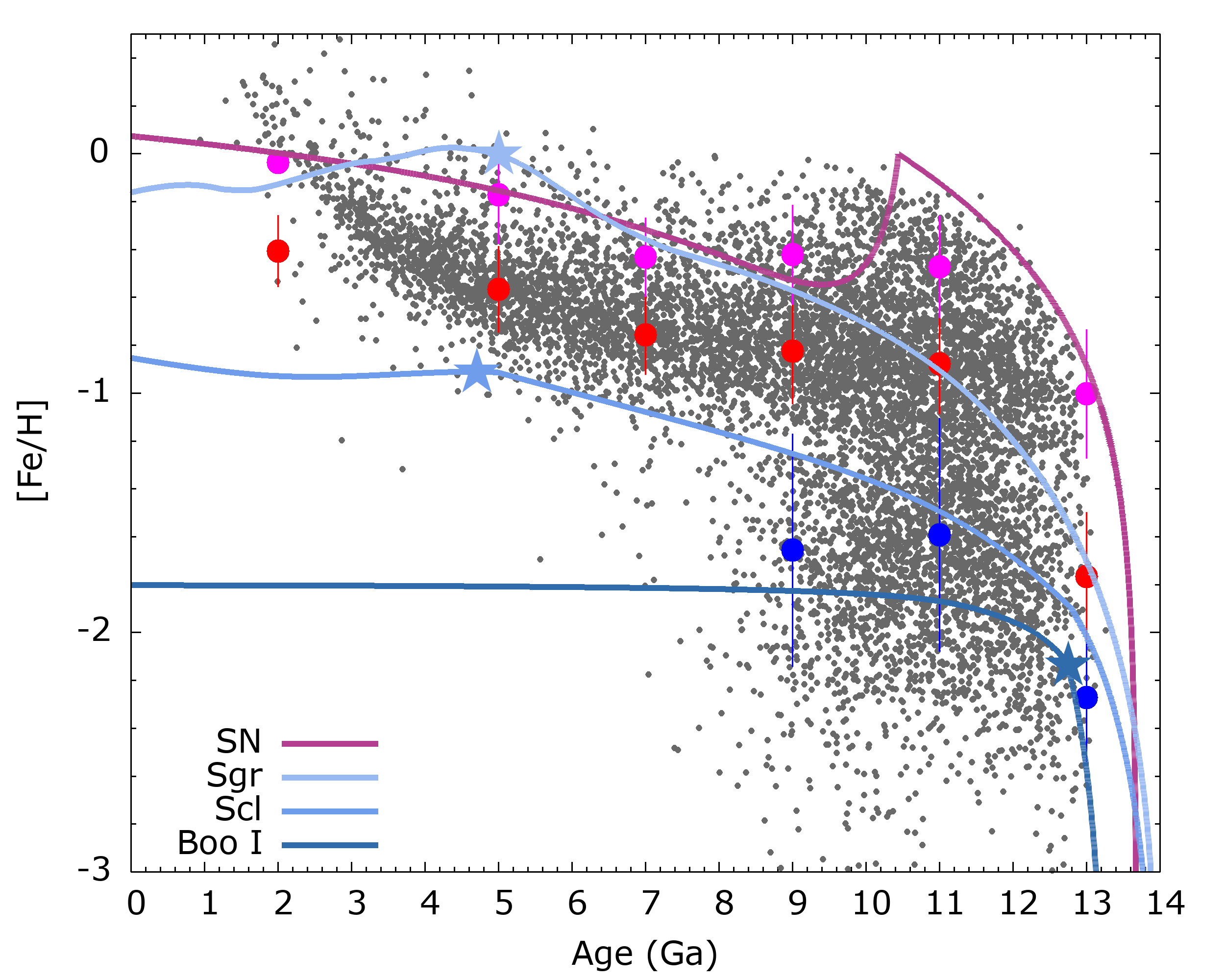}}
    \caption{AMRs predicted by chemical evolution models tailored to the solar neighbourhood (magenta line), the Sagittarius dwarf spheroidal galaxy (light blue line), the Sculptor dwarf spheroidal galaxy (medium blue line), and the Bo\"otes~I ultra-faint dwarf (dark blue line). The star symbols on top of each curve indicate the ages at which the star formation stops in the dwarf galaxy models. The theoretical AMRs are compared to the distribution of our good-quality TOPoS SG sample in the Age-[Fe/H] plane (grey dots and filled circles).}
    \label{fig:amr_GCE}
\end{figure}

In Fig.~\ref{fig:amr_GCE}, we present the AMRs for the solar neighbourhood (SN), the Sagittarius (Sgr) dwarf spheroidal galaxy, the Sculptor (Scl) dwarf spheroidal galaxy, and the Bo\"otes I (Boo~I) ultra-faint dwarf predicted by pure galactic chemical evolution models that have been presented and extensively discussed elsewhere  \citep{Romano2013,Romano2015,Mucciarelli2017,Spitoni2019,Spitoni2021}. The model predictions are compared with the relations defined from our dataset (see previous sections). The star symbols on the Sgr, Scl, and Boo~I tracks indicate the points at which star formation is truncated in the models.

The divergence in the theoretical tracks reflects the fundamental differences in star formation efficiency, gas accretion history, and the impact of galactic winds across varying mass scales. The Boo~I model \citep[model Boo~7 of][]{Romano2015} represents an ultra-faint dwarf galaxy with a stellar mass of 5.5~$\times 10^4$~M$_\odot$ and a stellar metallicity distribution peaking at [Fe/H]~$\simeq -$2.6 dex. The Scl model \citep[model Scl~T of][]{Romano2013} represents a dwarf galaxy with a stellar mass more than two orders of magnitude larger than that of the ultra-faint, and a stellar metallicity distribution peaking at [Fe/H]~$\simeq -$1.6 dex. The Sgr model aims at reproducing the main properties of the Sgr dwarf spheroidal galaxy at infall, namely, a rather massive object with a stellar mass of 7~$\times 10^8$~M$_\odot$ and a stellar metallicity distribution peak at around [Fe/H]~$\simeq -$0.6 dex.

The SN model (magenta line) exhibits the most rapid chemical enrichment, reaching solar metallicity within the first few billion years, when the in situ inner halo and thick-disc components form. A distinct feature of the model is the discontinuity at approximately 10.5~Ga ago. This `kink' is consistent with a dual-infall scenario, where a second episode of pristine gas accretion dilutes the interstellar medium before the onset of the thin-disc 
formation phase. 
The fact that the data are underpopulated in the region of the  kink does not necessarily signal a discrepancy between the data and
the model, as we know that our observational sample is biased against solar-metallicity stars \citep[see][]{toposVI}.
The Sgr model (light blue line) shows a protracted enrichment history. Although it eventually reaches near-solar metallicities around 5 Ga ago, its track remains systematically below the SN model for the majority of cosmic time. This is indicative of an SFR that is lower than the SN and, at the same time,  a potential well that is deeper
than that of the smaller dwarfs, allowing for continued enrichment despite periodic gas loss, mainly due to the interaction with the Milky Way in our model \citep[see][]{Mucciarelli2017}. 
{ 
In fact the model assumes a very massive dark matter halo for the Sgr progenitor ($6\times 10^{10}$ $\rm M_\odot$).
In the model the mass loss is tuned to reproduce the observed abundances and it must be 
understood as the sum of the mass loss driven by supernova events in the galaxy and tidal stripping from
the interaction with the Milky Way.}
In the low-mass regime, the Scl (medium blue) and Boo~I (dark blue) models demonstrate the effects of  galactic strangulation on chemical evolution. The Boo~I track reaches a metallicity plateau at [Fe/H]~$\approx -$1.8 after star formation halts, due to delayed Fe production from SNeIa.  This behaviour is typical of ultra- faint dwarfs, where early feedback from the first supernovae and the effects of reionisation effectively truncate star formation, yet leaving (in our models) abundant metal-poor gas associated with the galaxy when it infalls onto the Milky Way \citep[see][]{Romano2013,Romano2015}. This opens up the intriguing possibility that the high radial velocity, young, metal-poor stars recently found in the Galactic halo \citep[e.g.,][]{Bonifacio2024,Caffau2024} originated in dwarf satellites at their first interaction with the Milky Way \citep{Hammer2024}. 

The observations-model predictions comparison suggests that the observed metal-poor tail of the solar neighbourhood is not a single population, but a composite one formed in environments with low star formation efficiencies, similar to those found in dwarf satellites. The presence of these stars in the solar neighbourhood at various ages supports the hierarchical assembly of the Milky Way through the accretion of progenitors with varying chemical histories (see next section).
{  We stress that given the bias present in our observational sample, we are not attempting
a quantitative comparison between models and observations. Such a comparison requires an unbiased sample
or a good understanding of the selection function and should also take into account the phenomenon of radial
migration.}

\section{Comparison with cosmological models for the Local Group assembly}
\label{sec:cosmo}
In this Section, we compare our observational data with the cosmological chemical evolution model {\tt NEFERTITI} \citep[Near-FiEld cosmology: Re-Tracing Invisible Times;][]{Koutsouridou2023}, which is designed to study early galaxy formation processes and the nature of the first (Pop III) stars. The model follows the star formation, chemical enrichment, and assembly histories of galaxies across cosmic time, including the relevant physical and feedback processes. More specifically, it accounts for:
\begin{itemize}
\item incomplete sampling of the stellar initial mass function (IMF) in galaxies with low SFR \citep{Rossi2021};
\item the unknown properties of Pop III stars, which may have different IMFs \citep{Koutsouridou2024} and may explode as supernovae (SNe) with different energy distribution functions \citep{Koutsouridou2023};
\item the transition from Pop III to normal Pop II and Pop I stars \cite{Salvadori2010};
\item the production and injection of chemical elements, from C to Zn, by different stellar populations, also including SNIa and AGB stars \citep{Koutsouridou2025};
\item the inhomogeneous mixing of metals injected by SN-driven outflows into the surrounding intergalactic medium \cite[e.g.,][]{Salvadori2014, Koutsouridou2026};
\item the inhomogeneous process of reionisation \citep{Koutsouridou2026}.
\end{itemize}

We couple {\tt NEFERTITI} with a suite of $\sim$30 Caterpillar N-body simulations of the Local Group \citep{Griffen2016}, i.e., the highest-resolution dark-matter-only simulations currently available, which resolve minihalos that hosted the first Pop III stars \citep[e.g.,][]{Hirano2014} and are likely associated with present-day ultra-faint dwarf galaxies \citep[e.g.,][]{Salvadori2015}. The {\tt NEFERTITI} model not only reproduces the present-day  mass in stars and gas  as well as the metallicities of the Milky Way and the observed metallicity distribution function (MDF) of Galactic halo stars \citep[see][]{Koutsouridou2023}, but also matches the properties of Milky Way progenitors observed at $z\approx 8$ by JWST in the Firefly Sparkle system \citep{Rusta2024}. Therefore, it is an ideal model for studying the assembly of the Milky Way and uncovering the physical origin of the observed stellar AMR.

In Fig.~\ref{fig:nefertiti}, we compare the observed AMR in the Galactic halo with the predictions of {\tt NEFERTITI}. In the top panel, we show the {\it global AMR} for all present-day stars in the Milky Way and its galaxy satellites; in the middle panel, the AMR for stars in the Milky Way only; and in the bottom panel, the AMR for Milky Way stars that formed in situ, i.e., those that, at each epoch, formed exclusively within the most massive dark matter halo in each of the 30 merger histories.

Interestingly, the AMR for Milky Way stars (middle panel) matches the observational findings well. Note that, due to the lack of spatial information for present-day dark matter particles in the Caterpillar simulations, we cannot restrict our analysis to stars predicted to reside specifically in the Galactic stellar halo. As a result, the predicted relations are expected to be broader.

By comparing the different panels, it becomes clear that the low-metallicity AMR can be reproduced only when accreted stars are included in the Milky Way population, i.e., stars that originally formed in low-mass progenitor halos and were subsequently accreted by the Milky Way via merging processes. 
{  Although there is no kinematical or orbital characteristic that can unambiguously demonstrate that any
given star is the result of an accretion onto the Milky Way, we know that high eccentricity, low or negative angular momentum,
and large apocentric distances are clues that point towards accretion. 
In our sample the stars associated with the low-AMR (see Sect.\,\ref{sec:diff_amr}) 
display high eccentricities, 76\% of the sample have eccentricity larger than 0.5, the mean eccentricity
is 0.66, compared to 0.35 of the complementary sample. The mean angular momentum is   
  $\rm -138\, kpc\, km\,s^{-1}$, compared to $\rm 1292\, kpc\, km\,s^{-1}$ for the complementary sample.
  The mean apocentric distance is 14\,kpc, compared to 10\,kpc for the complementary sample.
  Although none of these facts ensures that any of the stars associated with the low-AMR has indeed been accreted,
  taken collectively, they strongly support the notion that a non-negligible fraction of them has been accreted.
}
For this reason, the AMR of these accreted stars is similar to that of stars in some of the Milky Way dwarf satellites. Overall, these results indicate that the most metal-poor and oldest stars in the Milky Way formed in dwarf galaxy progenitors, thus supporting the scenario proposed to explain the different AMRs discussed in Sec.~4.2.

\begin{figure}
    \centering
    \resizebox{0.5\textwidth}{!}{\includegraphics{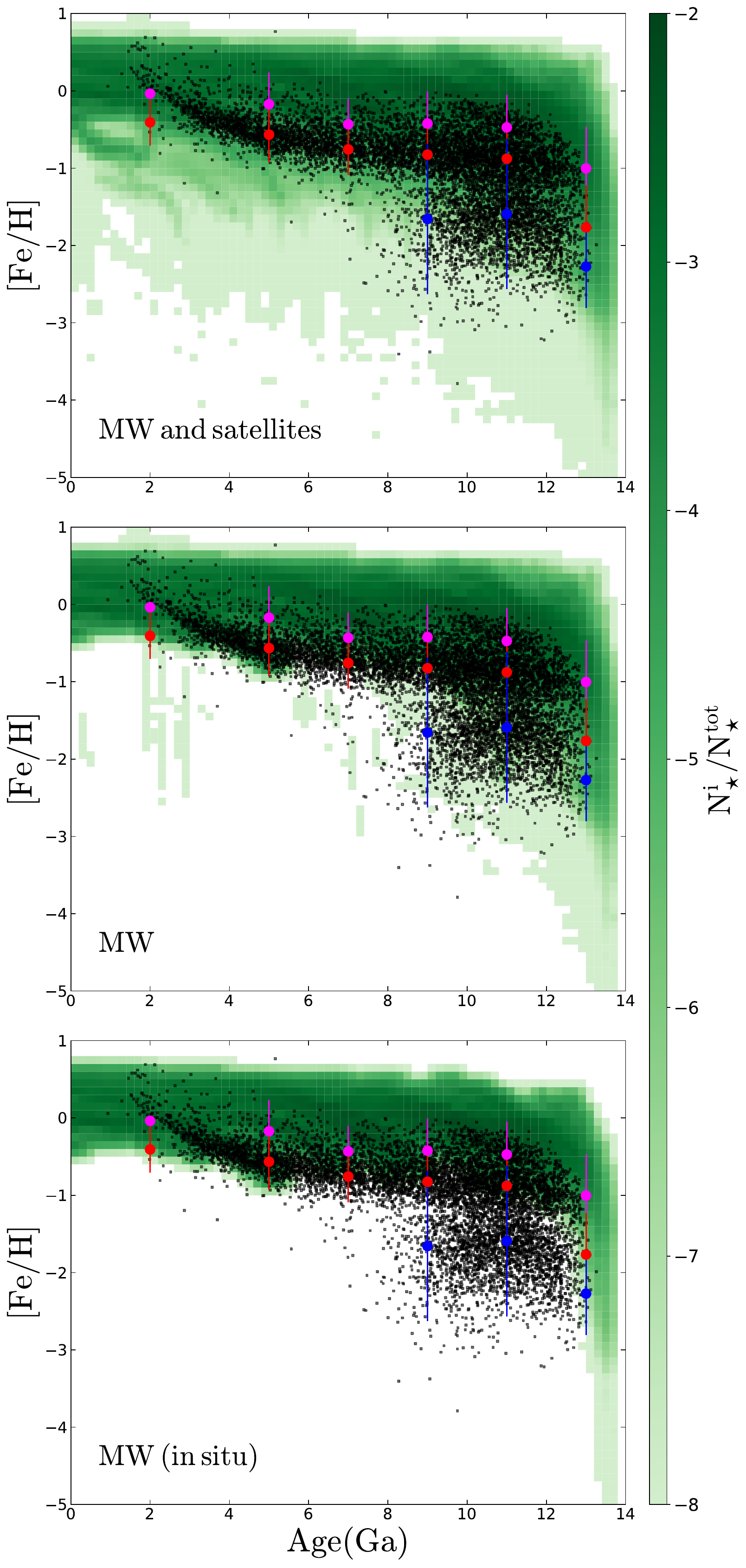}}
    \caption{Comparison between the data and the AMR predicted by the NEFERTITI model for stars residing in: the Milky Way and its dwarf satellites (top); the Milky Way, including stars formed both in situ and in accreted satellites (middle); and the Milky Way, considering only stars formed in situ (bottom).}
    \label{fig:nefertiti}
\end{figure}

\section{Discussion and conclusions}

In our sample of TOPoS SG stars we have been able to
highlight the existence of three distinct AMRs.
In our case, the metallicity of the stars is the fossil record of the
metallicity of the gas at the time they were formed. 
We tentatively associate the mid-AMR and high-AMR with the
Galactic discs. Although the thick disc stars mainly follow
the mid-AMR there is a non-negligible part that follows the high-AMR.
The reverse is true for thin disc stars. We therefore suggest that both
discs had some degree of heterogeneity.

The low-AMR can be loosely  associated with the Galactic halo, which includes
dynamical substructures such as GSE and Sequoia. The fact that a few halo
stars are found at metallicities above --1.0 can be understood  in terms
of the properties of the merging galaxies, which could host 
relatively metal-rich populations, 
and may have produced higher metallicity stars in the starburst triggered by the merger.

The fact that the low-AMR is only defined for the three oldest age bins and the large
dispersion in metallicities observed in these bins strongly suggest that in this epoch
the evolution of the Milky Way was dominated by mergers and that this was rather
abruptly terminated about 8\,Ga ago. The event changing the evolution
from merger-dominated to secular may have been the major merger associated with GSE,
but dedicated  simulations are necessary to confirm or refute this scenario. 

Our proposed scenario is, by and large, supported both by pure chemical evolution
models (Sec.\,\ref{sec:chem}) and by  cosmological models of the Local Group
(Sec.\,\ref{sec:cosmo}). We do not seek quantitative agreement between our data and the models, 
since, as demonstrated by \citet{toposVI}, our sample is extracted from a
sample that is highly biased in favour of metal-poor stars and against stars of solar metallicity or higher.
Although \citet{toposVI} corrected for bias for the purpose of the metallicity
distribution function, in our case it is not obvious how to do so, since we deal with a subset
that has  a further
selection of the precision of the parallaxes and the colours. 
We believe it is important that our analysis highlighted the role of mergers in the evolution
of the Milky Way up to 8~Ga ago. 
We are confident that large ongoing wide-field spectroscopic surveys such as WEAVE \citep{weave},
4MOST \citep{4most}, and Gaia \citep{gaia}, will provide larger and less biased samples
for which quantitative comparisons with models will be meaningful.
Furthermore, the ESA PLATO mission \citep{plato} will provide a large
sample of stars with asteroseismic ages, especially for giant stars, which will nicely complement the
evolutionary ages, based on SG stars, and will allow us to obtain a clearer view of the
AMR in the different components of the Milky Way.

\begin{acknowledgements}
We are grateful to the anonymous referee for the careful reading of the manuscript
and useful suggestions that helped to improve it.
We acknowledge the use of the Artificial Intelligence tool Emmy (\url{https://emmy.cnrs.fr})
provided by CNRS to perform part of the statistical analysis described in sections two and four.
DR acknowledges partial financial support from the
project \emph{``LEGO – Reconstructing the building blocks of the Galaxy by chemical tagging''} (PI A. Mucciarelli) granted by the Italian MUR through contract PRIN 2022LLP8TK\_001. I.K. acknowledges ERC support (grant agreement No. 101117455). L.M. gratefully acknowledges support from
ANID-FONDECYT Regular Project n. 1251809.
This work has made use of data from the European Space Agency (ESA) mission
{\it Gaia} (\url{https://www.cosmos.esa.int/gaia}), processed by the {\it Gaia}
Data Processing and Analysis Consortium (DPAC,
\url{https://www.cosmos.esa.int/web/gaia/dpac/consortium}). Funding for the DPAC
has been provided by national institutions, in particular the institutions
participating in the {\it Gaia} Multilateral Agreement.
This paper made use of data from Sloan Digital Sky Survey
\url{www.sdss.org}.
\end{acknowledgements}

   \bibliographystyle{aa} % style aa.bst

   \bibliography{aa60236-26corr} % your refere

\appendix

\section{Estimating luminosities and associated errors}
\label{ap_lum}

{The relations between surface gravity $g$ of a star of mass $M$ and radius $R$, its luminosity $L$
and effective temperature $T_\mathrm{eff}$ are summarised in equations A.1 -- A.3.} 

\begin{equation}
    {g\over g_\odot} = \left({M\over M_\odot}\right)\times {\left(R\over R_\odot\right)^{-2}}
\end{equation}

\begin{equation}
    \left({R\over R_\odot}\right)^2 = \left({L\over L_\odot}\right)\times \left({T_\mathrm{eff}\over T_\mathrm{eff, \odot}}\right)^{-4}
\end{equation}

\begin{equation}
    {g\over g_\odot} =  \left({M\over M_\odot}\right)\times\left({L\over L_\odot}\right)^{-1} \times \left({T_\mathrm{eff}\over T_\mathrm{eff, \odot}}\right)^4
\end{equation}

In the sample of \citet{toposVI} the surface gravity is
provided, this has been obtained assuming a stellar mass of 0.8M$_\odot$. 
{  To determine ages we prefer to
use the surface luminosity. }
In order to derive the stellar luminosity we re-arrange equation
A.3 to yield:

\begin{equation}
    \log \left({L\over L_\odot}\right) =
     \left( \log g_\odot - \log g\right)+\log\left({M\over M_\odot}\right) +4\log\left({T_\mathrm{eff}\over T_\mathrm{eff, \odot}}\right)
\end{equation}
{  
We used this equation assuming a mass of 0.8M$_\odot$, that is consistent
with the log g provided in \citep{toposVI}.
}
Taking into account that the relation between luminosities
and bolometric magnitudes is

\begin{equation}
    \left(\mu_\mathrm{bol}-\mu_\mathrm{bol\odot}\right) =
    -2.5\left({\log\left(L\over L_\odot\right)}\right)
\end{equation}

Where $\mu$ denotes absolute magnitudes, and recalling that
the absolute bolometric magnitude of the Sun is 4.74 
\citep{mamajek2015iau2015resolutionb2} and introducing the parallax
of the star $\varpi$, and recalling that 

\begin{equation}
\mu_\mathrm{bol} = (G+G_c) +5 +5 \log\left(\varpi\right) 
\end{equation}

where $G$ is the Gaia $G$ magnitude and
$G_c$ is the corresponding bolometric correction,
one can also write

\begin{equation}
    \log\left({L\over L_\odot}\right)
    = -0.4(G+G_c)  +(0.4\times 4.74 -2) -2\log\left(\varpi\right) \label{l2}
\end{equation}

Equation \ref{l2} 
is useful because it allows to establish that
error on the luminosity depends only on 
the error on $G$ and on the parallax

in practice

\begin{equation}
    \Delta\log\left({L\over L_\odot}\right) =
    0.4\Delta G + {2\over \ln\left(10\right)}{\Delta\varpi\over \varpi} \label{l_err}
\end{equation}
We used equation A.4 to compute $\log \left({L\over L_\odot}\right)$
for all the stars in the ``good parallax'' sample of
\citet{toposVI}. 
Equation A.8 was used to estimate the errors on logarithmic
luminosity. As error in $G$ magnitude we used the estimate
provided by CDS, that besides the photon-noise error
includes a {\em floor} of $G_f = 0.0027553202$\,mag\footnote{See
\url{https://cdsarc.cds.unistra.fr/viz-bin/cat/I/350}.},
in practice:
\begin{equation}
    \Delta G = \sqrt{\left[\left(-2.5\over \ln{\left(10\right)}\right)
    \left({\Delta\left(\mathrm{Flux}\right)\over \mathrm{Flux}}\right)\right]^2 +  G_f^2}
\end{equation}

\section{Using ages derived without any prior }
\label{noprior}

\begin{figure}
    \centering
    \resizebox{0.9\hsize}{!}{\includegraphics{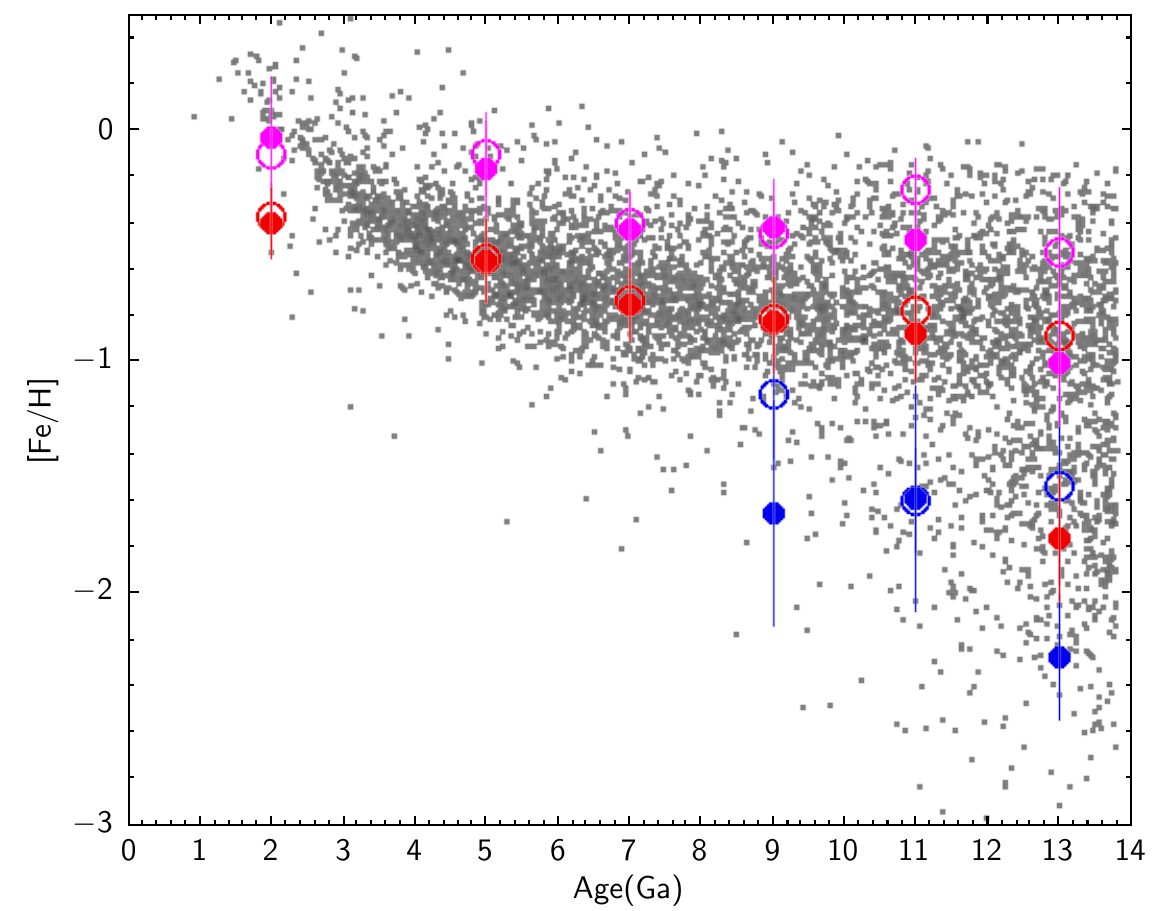}}
    \caption{Same Fig.\,\ref{agemet_peaks}, except that the ages
     are derived without any prior, the open circles  identify the peaks in any age bin, 
     while the filled circles show the peaks derived from the dataset where ages
     were derived using a flat prior imposing ages to be  in the interval from zero to 13.8\,Ga.
     The colour coding is the same as in Fig.\,\ref{agemet_peaks}.}
    \label{agemet_noprior}
\end{figure}

In Fig.\,\ref{agemet_noprior} we show the age-metallicity relation
using the ages derived without any prior, we shall refer to this sample as the ``no-prior'' sample. We retained only the stars
for which SPInS derived an age  smaller than the age of the Universe,
this reduces the sample to 60\% of the original sample (5\,273 stars).
We analysed the no-prior sample in the same way as described in Sect.\ref{sec:diff_amr},
the position of the peaks is denoted by open circles, while the filled circles
are the position of the peaks in Fig.\,\ref{agemet_peaks}. 
The most noticeable differences are in the oldest age bin,
where the peaks in the no-prior sample appear systematically at higher
metallicities even by 1\,dex. The differences become smaller and smaller
as we move to the younger metallicity bins. We conclude that the general morphology
of the age-metallicity relations does not depend on whether one applies a prior on age
or not in the age determination. We stress that our sample, whatever the hypothesis
made on the age determination, is biased in favour of metal-poor stars as discussed in
\citet{toposVI}. In order to determine precisely the position of the peaks, especially
in the oldest age bins, we need an unbiased sample.

\end{document}